\documentclass[
	aps,
	prb,
	twocolumn,
	amssymb,
	amsmath,
	notitlepage,
	floatfix,
	nofootinbib,
	a4paper,
	citeautoscript,
	superscriptaddress
]{revtex4-1}

\usepackage[usenames,dvipsnames]{xcolor}
  \definecolor{goodgreen}{rgb}{0.1,0.5,0}  
\usepackage{hyperref}
\hypersetup{
    colorlinks=true,
    linkcolor=blue,
    filecolor=black,
    urlcolor=goodgreen,
    citecolor=blue
}
\usepackage{braket}
\usepackage{multirow}
\usepackage{setspace}
\newcommand*\bs[1]{\textbf{\textsf{#1}}} 
\newcommand{\sect}[1]{{\noindent\textbf{#1.}~}\ignorespaces}
\let\oldcaption\caption
\renewcommand{\caption}[2][]{
    \oldcaption[#1]{\bs{#1.} #2}
}

\usepackage{tikz}
\usetikzlibrary{arrows}
\usetikzlibrary{arrows.meta}
\usetikzlibrary{backgrounds}
\usetikzlibrary{decorations.markings}
\usepackage{xstring}

\newcommand{\sarrow}[3]{\draw[#1,>=latex,NavyBlue,line width=2] (#2) -- (#3);}
\newcommand{\sx}[6]{%
	\vcenter{\hbox{\scalebox{0.65}{
		\begin{tikzpicture}
			\path[use as bounding box] (0.5,0.35) rectangle (2.5,2.65);
			
			\def \dy {0.75}
			\def \ll {0.7}			
			\def \amin {-0.4}		
			\def \amax {0.5}		
			\def \dxspins {0.2}		
			\def \dyspins {-0.1}	
			
			\node (001us) at (1,\dy+\amin) {};
			\node (001ue) at (1,\dy+\amax) {};		
			\node (001ds) at (1,\dy+\amin+\dyspins) {};
			\node (001de) at (1,\dy+\amax+\dyspins) {};
			
			\node (002us) at (1+\dxspins/2,\dy+\amin) {};
			\node (002ue) at (1+\dxspins/2,\dy+\amax) {};
			\node (002ds) at (1-\dxspins/2,\dy+\amin+\dyspins) {};
			\node (002de) at (1-\dxspins/2,\dy+\amax+\dyspins) {};
			
			\node (011us) at (2,\dy+\amin) {};
			\node (011ue) at (2,\dy+\amax) {};		
			\node (011ds) at (2,\dy+\amin+\dyspins) {};
			\node (011de) at (2,\dy+\amax+\dyspins) {};
			
			\node (012us) at (2+\dxspins/2,\dy+\amin) {};
			\node (012ue) at (2+\dxspins/2,\dy+\amax) {};
			\node (012ds) at (2-\dxspins/2,\dy+\amin+\dyspins) {};
			\node (012de) at (2-\dxspins/2,\dy+\amax+\dyspins) {};
			
			\node (101us) at (1,2*\dy+\amin) {};
			\node (101ue) at (1,2*\dy+\amax) {};		
			\node (101ds) at (1,2*\dy+\amin+\dyspins) {};
			\node (101de) at (1,2*\dy+\amax+\dyspins) {};
			
			\node (102us) at (1+\dxspins/2,2*\dy+\amin) {};
			\node (102ue) at (1+\dxspins/2,2*\dy+\amax) {};
			\node (102ds) at (1-\dxspins/2,2*\dy+\amin+\dyspins) {};
			\node (102de) at (1-\dxspins/2,2*\dy+\amax+\dyspins) {};
			
			\node (111us) at (2,2*\dy+\amin) {};
			\node (111ue) at (2,2*\dy+\amax) {};		
			\node (111ds) at (2,2*\dy+\amin+\dyspins) {};
			\node (111de) at (2,2*\dy+\amax+\dyspins) {};
			
			\node (112us) at (2+\dxspins/2,2*\dy+\amin) {};
			\node (112ue) at (2+\dxspins/2,2*\dy+\amax) {};
			\node (112ds) at (2-\dxspins/2,2*\dy+\amin+\dyspins) {};
			\node (112de) at (2-\dxspins/2,2*\dy+\amax+\dyspins) {};
			
			\node (201us) at (1,3*\dy+\amin) {};
			\node (201ue) at (1,3*\dy+\amax) {};		
			\node (201ds) at (1,3*\dy+\amin+\dyspins) {};
			\node (201de) at (1,3*\dy+\amax+\dyspins) {};
			
			\node (202us) at (1+\dxspins/2,3*\dy+\amin) {};
			\node (202ue) at (1+\dxspins/2,3*\dy+\amax) {};
			\node (202ds) at (1-\dxspins/2,3*\dy+\amin+\dyspins) {};
			\node (202de) at (1-\dxspins/2,3*\dy+\amax+\dyspins) {};
			
			\node (211us) at (2,3*\dy+\amin) {};
			\node (211ue) at (2,3*\dy+\amax) {};		
			\node (211ds) at (2,3*\dy+\amin+\dyspins) {};
			\node (211de) at (2,3*\dy+\amax+\dyspins) {};
			
			\node (212us) at (2+\dxspins/2,3*\dy+\amin) {};
			\node (212ue) at (2+\dxspins/2,3*\dy+\amax) {};
			\node (212ds) at (2-\dxspins/2,3*\dy+\amin+\dyspins) {};
			\node (212de) at (2-\dxspins/2,3*\dy+\amax+\dyspins) {};
			
			\draw[line width = 1] (1-\ll/2,\dy) -- (1+\ll/2,\dy);
			\draw[line width = 1] (2-\ll/2,\dy) -- (2+\ll/2,\dy);
			
			\draw[line width = 1] (1-\ll/2,2*\dy) -- (1+\ll/2,2*\dy);
			\draw[line width = 1] (2-\ll/2,2*\dy) -- (2+\ll/2,2*\dy);
			
			\draw[line width = 1] (1-\ll/2,3*\dy) -- (1+\ll/2,3*\dy);
			\draw[line width = 1] (2-\ll/2,3*\dy) -- (2+\ll/2,3*\dy);
			
			\IfEqCase{#1}{%
				{0}{}
				{1}{\sarrow{->}{001us}{001ue}}
				{u}{\sarrow{->}{001us}{001ue}}
				{-1}{\sarrow{<-}{001ds}{001de}}
				{d}{\sarrow{<-}{001ds}{001de}}
				{2}{\sarrow{<-}{002ds}{002de}\sarrow{->}{002us}{002ue}}
			}[\PackageError{sx}{Undefined option to: #1}{}]%
			\IfEqCase{#2}{%
				{0}{}
				{1}{\sarrow{->}{011us}{011ue}}
				{u}{\sarrow{->}{011us}{011ue}}
				{-1}{\sarrow{<-}{011ds}{011de}}
				{d}{\sarrow{<-}{011ds}{011de}}
				{2}{\sarrow{<-}{012ds}{012de}\sarrow{->}{012us}{012ue}}
			}[\PackageError{sx}{Undefined option to: #2}{}]%
			
			\IfEqCase{#3}{%
				{0}{}
				{1}{\sarrow{->}{101us}{101ue}}
				{u}{\sarrow{->}{101us}{101ue}}
				{-1}{\sarrow{<-}{101ds}{101de}}
				{d}{\sarrow{<-}{101ds}{101de}}
				{2}{\sarrow{<-}{102ds}{102de}\sarrow{->}{102us}{102ue}}
			}[\PackageError{sx}{Undefined option to: #3}{}]%
			\IfEqCase{#4}{%
				{0}{}
				{1}{\sarrow{->}{111us}{111ue}}
				{u}{\sarrow{->}{111us}{111ue}}
				{-1}{\sarrow{<-}{111ds}{111de}}
				{d}{\sarrow{<-}{111ds}{111de}}
				{2}{\sarrow{<-}{112ds}{112de}\sarrow{->}{112us}{112ue}}
			}[\PackageError{sx}{Undefined option to: #4}{}]%
			
			\IfEqCase{#5}{%
				{0}{}
				{1}{\sarrow{->}{201us}{201ue}}
				{u}{\sarrow{->}{201us}{201ue}}
				{-1}{\sarrow{<-}{201ds}{201de}}
				{d}{\sarrow{<-}{201ds}{201de}}
				{2}{\sarrow{<-}{202ds}{202de}\sarrow{->}{202us}{202ue}}
			}[\PackageError{sx}{Undefined option to: #5}{}]%
			\IfEqCase{#6}{%
				{0}{}
				{1}{\sarrow{->}{211us}{211ue}}
				{u}{\sarrow{->}{211us}{211ue}}
				{-1}{\sarrow{<-}{211ds}{211de}}
				{d}{\sarrow{<-}{211ds}{211de}}
				{2}{\sarrow{<-}{212ds}{212de}\sarrow{->}{212us}{212ue}}
			}[\PackageError{sx}{Undefined option to: #6}{}]%
			
		\end{tikzpicture}
	}}}
}

\newcommand{\ssarrow}[3]{\draw[#1,>={Latex[length=7, width=8]},NavyBlue,line width=2.5] (#2) -- (#3);}
\newcommand{\ssx}[6]{%
	\vcenter{\hbox{\scalebox{0.4}{
		\begin{tikzpicture}			
			\path[use as bounding box] (0.5,0.2) rectangle (2.4,2.3);
			
			\def \dy {0.6}
			\def \ll {0.7}			
			\def \amin {-0.4}		
			\def \amax {0.5}		
			\def \dxspins {0.25}		
			\def \dyspins {-0.1}	
			
			\node (001us) at (1,\dy+\amin) {};
			\node (001ue) at (1,\dy+\amax) {};		
			\node (001ds) at (1,\dy+\amin+\dyspins) {};
			\node (001de) at (1,\dy+\amax+\dyspins) {};
			
			\node (002us) at (1+\dxspins/2,\dy+\amin) {};
			\node (002ue) at (1+\dxspins/2,\dy+\amax) {};
			\node (002ds) at (1-\dxspins/2,\dy+\amin+\dyspins) {};
			\node (002de) at (1-\dxspins/2,\dy+\amax+\dyspins) {};
			
			\node (011us) at (2,\dy+\amin) {};
			\node (011ue) at (2,\dy+\amax) {};		
			\node (011ds) at (2,\dy+\amin+\dyspins) {};
			\node (011de) at (2,\dy+\amax+\dyspins) {};
			
			\node (012us) at (2+\dxspins/2,\dy+\amin) {};
			\node (012ue) at (2+\dxspins/2,\dy+\amax) {};
			\node (012ds) at (2-\dxspins/2,\dy+\amin+\dyspins) {};
			\node (012de) at (2-\dxspins/2,\dy+\amax+\dyspins) {};
			
			\node (101us) at (1,2*\dy+\amin) {};
			\node (101ue) at (1,2*\dy+\amax) {};		
			\node (101ds) at (1,2*\dy+\amin+\dyspins) {};
			\node (101de) at (1,2*\dy+\amax+\dyspins) {};
			
			\node (102us) at (1+\dxspins/2,2*\dy+\amin) {};
			\node (102ue) at (1+\dxspins/2,2*\dy+\amax) {};
			\node (102ds) at (1-\dxspins/2,2*\dy+\amin+\dyspins) {};
			\node (102de) at (1-\dxspins/2,2*\dy+\amax+\dyspins) {};
			
			\node (111us) at (2,2*\dy+\amin) {};
			\node (111ue) at (2,2*\dy+\amax) {};		
			\node (111ds) at (2,2*\dy+\amin+\dyspins) {};
			\node (111de) at (2,2*\dy+\amax+\dyspins) {};
			
			\node (112us) at (2+\dxspins/2,2*\dy+\amin) {};
			\node (112ue) at (2+\dxspins/2,2*\dy+\amax) {};
			\node (112ds) at (2-\dxspins/2,2*\dy+\amin+\dyspins) {};
			\node (112de) at (2-\dxspins/2,2*\dy+\amax+\dyspins) {};
			
			\node (201us) at (1,3*\dy+\amin) {};
			\node (201ue) at (1,3*\dy+\amax) {};		
			\node (201ds) at (1,3*\dy+\amin+\dyspins) {};
			\node (201de) at (1,3*\dy+\amax+\dyspins) {};
			
			\node (202us) at (1+\dxspins/2,3*\dy+\amin) {};
			\node (202ue) at (1+\dxspins/2,3*\dy+\amax) {};
			\node (202ds) at (1-\dxspins/2,3*\dy+\amin+\dyspins) {};
			\node (202de) at (1-\dxspins/2,3*\dy+\amax+\dyspins) {};
			
			\node (211us) at (2,3*\dy+\amin) {};
			\node (211ue) at (2,3*\dy+\amax) {};		
			\node (211ds) at (2,3*\dy+\amin+\dyspins) {};
			\node (211de) at (2,3*\dy+\amax+\dyspins) {};
			
			\node (212us) at (2+\dxspins/2,3*\dy+\amin) {};
			\node (212ue) at (2+\dxspins/2,3*\dy+\amax) {};
			\node (212ds) at (2-\dxspins/2,3*\dy+\amin+\dyspins) {};
			\node (212de) at (2-\dxspins/2,3*\dy+\amax+\dyspins) {};
			
			\draw[line width = 1] (1-\ll/2,\dy) -- (1+\ll/2,\dy);
			\draw[line width = 1] (2-\ll/2,\dy) -- (2+\ll/2,\dy);
			
			\draw[line width = 1] (1-\ll/2,2*\dy) -- (1+\ll/2,2*\dy);
			\draw[line width = 1] (2-\ll/2,2*\dy) -- (2+\ll/2,2*\dy);
			
			\draw[line width = 1] (1-\ll/2,3*\dy) -- (1+\ll/2,3*\dy);
			\draw[line width = 1] (2-\ll/2,3*\dy) -- (2+\ll/2,3*\dy);
			
			\IfEqCase{#1}{%
				{0}{}
				{1}{\ssarrow{->}{001us}{001ue}}
				{u}{\ssarrow{->}{001us}{001ue}}
				{-1}{\ssarrow{<-}{001ds}{001de}}
				{d}{\ssarrow{<-}{001ds}{001de}}
				{2}{\ssarrow{<-}{002ds}{002de}\ssarrow{->}{002us}{002ue}}
			}[\PackageError{sx}{Undefined option to: #1}{}]%
			\IfEqCase{#2}{%
				{0}{}
				{1}{\ssarrow{->}{011us}{011ue}}
				{u}{\ssarrow{->}{011us}{011ue}}
				{-1}{\ssarrow{<-}{011ds}{011de}}
				{d}{\ssarrow{<-}{011ds}{011de}}
				{2}{\ssarrow{<-}{012ds}{012de}\ssarrow{->}{012us}{012ue}}
			}[\PackageError{sx}{Undefined option to: #2}{}]%
			
			\IfEqCase{#3}{%
				{0}{}
				{1}{\ssarrow{->}{101us}{101ue}}
				{u}{\ssarrow{->}{101us}{101ue}}
				{-1}{\ssarrow{<-}{101ds}{101de}}
				{d}{\ssarrow{<-}{101ds}{101de}}
				{2}{\ssarrow{<-}{102ds}{102de}\ssarrow{->}{102us}{102ue}}
			}[\PackageError{sx}{Undefined option to: #3}{}]%
			\IfEqCase{#4}{%
				{0}{}
				{1}{\ssarrow{->}{111us}{111ue}}
				{u}{\ssarrow{->}{111us}{111ue}}
				{-1}{\ssarrow{<-}{111ds}{111de}}
				{d}{\ssarrow{<-}{111ds}{111de}}
				{2}{\ssarrow{<-}{112ds}{112de}\ssarrow{->}{112us}{112ue}}
			}[\PackageError{sx}{Undefined option to: #4}{}]%
			
			\IfEqCase{#5}{%
				{0}{}
				{1}{\ssarrow{->}{201us}{201ue}}
				{u}{\ssarrow{->}{201us}{201ue}}
				{-1}{\ssarrow{<-}{201ds}{201de}}
				{d}{\ssarrow{<-}{201ds}{201de}}
				{2}{\ssarrow{<-}{202ds}{202de}\ssarrow{->}{202us}{202ue}}
			}[\PackageError{sx}{Undefined option to: #5}{}]%
			\IfEqCase{#6}{%
				{0}{}
				{1}{\ssarrow{->}{211us}{211ue}}
				{u}{\ssarrow{->}{211us}{211ue}}
				{-1}{\ssarrow{<-}{211ds}{211de}}
				{d}{\ssarrow{<-}{211ds}{211de}}
				{2}{\ssarrow{<-}{212ds}{212de}\ssarrow{->}{212us}{212ue}}
			}[\PackageError{sx}{Undefined option to: #6}{}]%
			
		\end{tikzpicture}
	}}}
}

\makeatletter
\renewcommand{\fnum@figure}{\bs{Fig.~\thefigure}}
\renewcommand{\fnum@table}{\bs{Tab.~\thetable}}
\makeatother

\renewcommand*\vec[1]{\ensuremath{\boldsymbol{#1}}}

\usepackage{graphicx}
\usepackage{soul}

\begin{document}

\title{Dark states in a carbon nanotube quantum dot}

\author{Andrea Donarini}
\affiliation{Institute for Theoretical Physics, University of Regensburg, 93040 Regensburg, Germany}

\author{Michael Niklas}
\altaffiliation{These two authors contributed equally}
\affiliation{Institute for Theoretical Physics, University of Regensburg, 93040 Regensburg, Germany}

\author{Michael Schafberger}
\altaffiliation{These two authors contributed equally}
\affiliation{Institute for Experimental and Applied Physics, University of Regensburg, 93040 Regensburg, Germany}

\author{Nicola Paradiso}
\affiliation{Institute for Experimental and Applied Physics, University of Regensburg, 93040 Regensburg, Germany}

\author{Christoph Strunk}
\altaffiliation{E-mail: \href{mailto:christoph.strunk@ur.de}{christoph.strunk@ur.de}}
\affiliation{Institute for Experimental and Applied Physics, University of Regensburg, 93040 Regensburg, Germany}

\author{Milena Grifoni}
\altaffiliation{E-mail: \href{mailto:milena.grifoni@ur.de}{milena.grifoni@ur.de}}
\affiliation{Institute for Theoretical Physics, University of Regensburg, 93040 Regensburg, Germany}

\date{\today}

\pacs{%
73.21.La,   
73.23.Hk,	
73.40.Gk,	
73.63.Fg	
}

\begin{abstract}
Illumination of atoms by resonant lasers can pump electrons into a coherent superposition of hyperfine levels which can no longer absorb the light. Such superposition is known as  dark state,  because fluorescent light emission is then suppressed. Here we report an all-electric analogue of this destructive interference effect in a carbon nanotube quantum dot. The dark states are a coherent superposition of valley (angular momentum) states which are  decoupled from either the drain or the source leads.  
Their emergence is visible in asymmetric current-voltage characteristics, with missing current steps and current suppression which depend on the polarity of the applied source-drain bias.  Our results demonstrate for the first time coherent-population trapping by all-electric means in an artificial atom. 
\end{abstract}

\maketitle
Coherent population trapping (CPT) occurs in three-level, Lambda-type atomic systems coupled to two quasi-resonant electromagnetic modes \citep{Arimondo1976, Whitley1976}. The light beams coherently excite the atom from two low-lying states to a common excited state. By proper detuning of the lasers, the system has a finite probability to decay into a coherent superposition of the low-lying states which  is decoupled from the light, a so called dark state (DS). In the stationary limit, the DS is occupied with probability one and the coherent population trapping is perfect.  
%
%
%
%

Quantum dots offer the possibility to engineer artificial atoms and molecules by proper circuit design, and hence to probe CPT in effective Lambda-systems. 
Early proposals \citep{Brandes2000, Brandes2005, Sanchez2008} have considered microwave irradiated  double quantum dot analogs of the seminal experiment \citep{Alzetta1976}. Since localization of the electrons in the DS also implies a vanishing current through the double quantum dot, this allows the electrical detection of CPT by recording variations of the current as the microwaves parameters are tuned.  All-electrical realizations of CPT have been proposed for triple quantum dot systems \citep{Michaelis2006, Emary2007, Poeltl2009, Kostyrko2009, Donarini2009, Busl2010, Sanchez2014, Niklas2017} and single molecules contacted in meta configuration \citep{ Begemann2008, Darau2009}. 
Despite the large number of theoretical proposals, the experimental observation of CPT in quantum dot setups has remained elusive so far. 

\begin{figure*}[t!]
	\centering
	\includegraphics{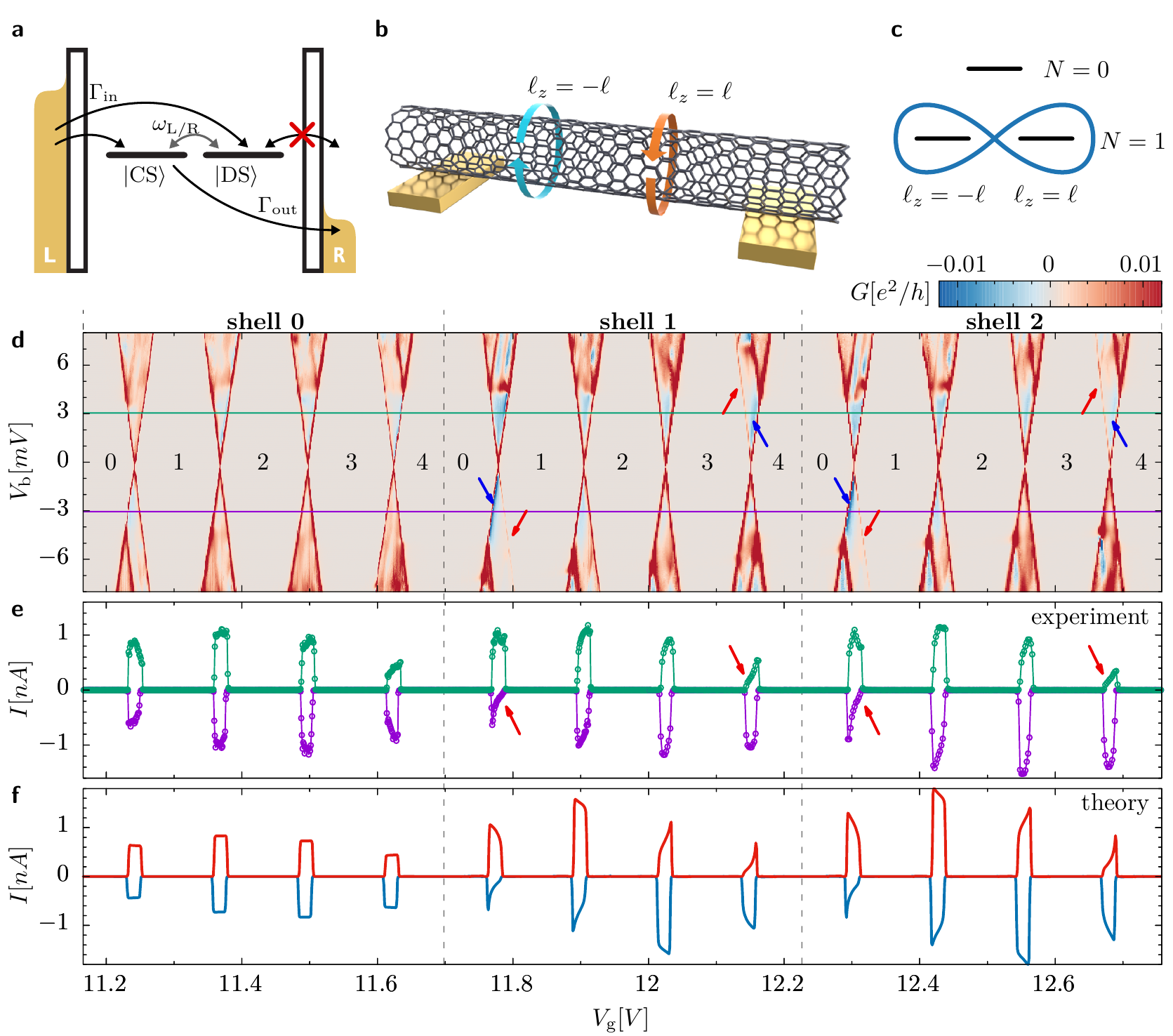}
	\caption[All-electronic dark states in transport through a carbon nanotube (CNT) quantum dot]{
		\bs{a}
		Quantum dot set-up to probe coherent population trapping in a dark state (DS). For the chosen bias voltage polarity, an electron can enter a DS from the left lead but it can not leave it by tunneling to the right lead.  Precession allows population transfer from the dark state to the coupled state (CS).   
		\bs{b}
		A CNT features degenerate angular momentum states. Linear combinations of angular momentum states for single-electron occupancy $(N=1)$ of a given nanotube shell, give rise to CS and DS. The latter is represented by the eight-shaped orbital in panel (\bs{c}).
		\bs{c}		
		Effective Lambda-system for coherent population trapping in a CNT quantum dot. 
		Because an electron in the DS  cannot leave the carbon nanotube, the state $N=0$, corresponding to an empty shell configuration, can no longer be reached.
		\bs{d}
		Experimental differential conductance as function of back-gate voltage $V_\mathrm{g}$ and bias voltage $V_\mathrm{b}$. Twelve consecutive Coulomb diamonds can be assigned to three CNT shells which get progressively occupied with $N=1$ to $N=4$ electrons. Current suppression (negative differential conductance) and faint Coulomb diamond borders are observed, indicated by blue and red arrows, respectively.  
		\bs{e}
		Current vs gate voltage for the two values  $V_\mathrm{b}=\pm 3.045$ mV of the bias voltages corresponding to the green/purple lines in panel (\bs{d}). Current suppression associated to coherent population trapping is indicated by red arrows.
		\bs{f}
		Numerically evaluated stationary current qualitatively reproducing the experiment.
		The parameters used in the simulation are in Tab.~II of the Methods.	
		}
	\label{fig1}
\end{figure*}

In this work we report the all-electrical realization of CPT  in a  single  carbon nanotube (CNT) quantum dot. As discussed below, the effect is quite generic, and requires the presence of  orbitally degenerate states which can form coherent superpositions. Under given conditions, tunneling events into and out of the dot successively trap the system in a DS, i.e., a coherent superposition of the degenerate levels which is decoupled from one of the leads. This in turn yields a characteristic current suppression as the bias voltage or the gate voltage are tuned.
%
%
%
 The situation is illustrated in the quantum dot setup of Fig.~\ref{fig1}(a) for the case of  a positive electrochemical potential drop between left and right leads. The coherent superposition of two degenerate states results in a coupled state (CS) and a DS which is decoupled from the right lead.
%
 %
 %
This allows electrons to enter the DS from the left while preventing them to leave it to any of the two leads. CPT occurs and current is suppressed. For opposite bias no suppression takes place. 
  %
In this work we demonstrate that such a situation has been realized in a CNT-based quantum dot.

 Similar to graphene, CNTs posses an orbital (valley) degree of freedom, arising from the two inequivalent Dirac points $K,K^\prime$ in the  Brillouin zone. In CNTs of the zig-zag class, such  orbital degree of freedom  corresponds to the longitudinal orbital momentum $\hbar \ell_z$, accounting for clockwise  ($\ell_z=-\ell$) or anti-clockwise ($\ell_z=\ell$) rotations along the tube waist \cite{Laird2015}, see Fig.~\ref{fig1}(b).
The finite length of the CNT quantum dot results in discrete longitudinal momentum states, with quantum number $m$, and hence in a shell structure like for the atomic bound states.
In the absence of spin-orbit coupling 
 \citep{Ando2000, Kuemmeth2008, Steele2013} and valley mixing \citep{Kuemmeth2008, Jespersen2011, Izumida2015, Marganska2015}, each shell consists of four degenerate bound states with spin, $\sigma =  \,\uparrow, \downarrow$, and angular momentum, $ \ell_z = \pm  \ell$, degrees of freedom. Thus each shell can accommodate up to $N=4$ electrons. As it will be shown in the following, for a CNT quantum dot near the $N=0 \leftrightarrow N=1$ resonance, a coherent superposition of angular momentum states can form which is decoupled from one of the two leads, and hence is a DS for the quantum dot for appropriate polarity of the applied bias voltage, see Fig.~\ref{fig1}(c). Due to the particle-hole symmetry of the many-body spectrum,  CPT  is expected near the $N=3 \leftrightarrow N=4$ resonance if the voltage polarity is reversed. 
 
\section*{Experimental signatures of CPT}
\begin{figure}
	\centering
	\includegraphics[width=88mm]{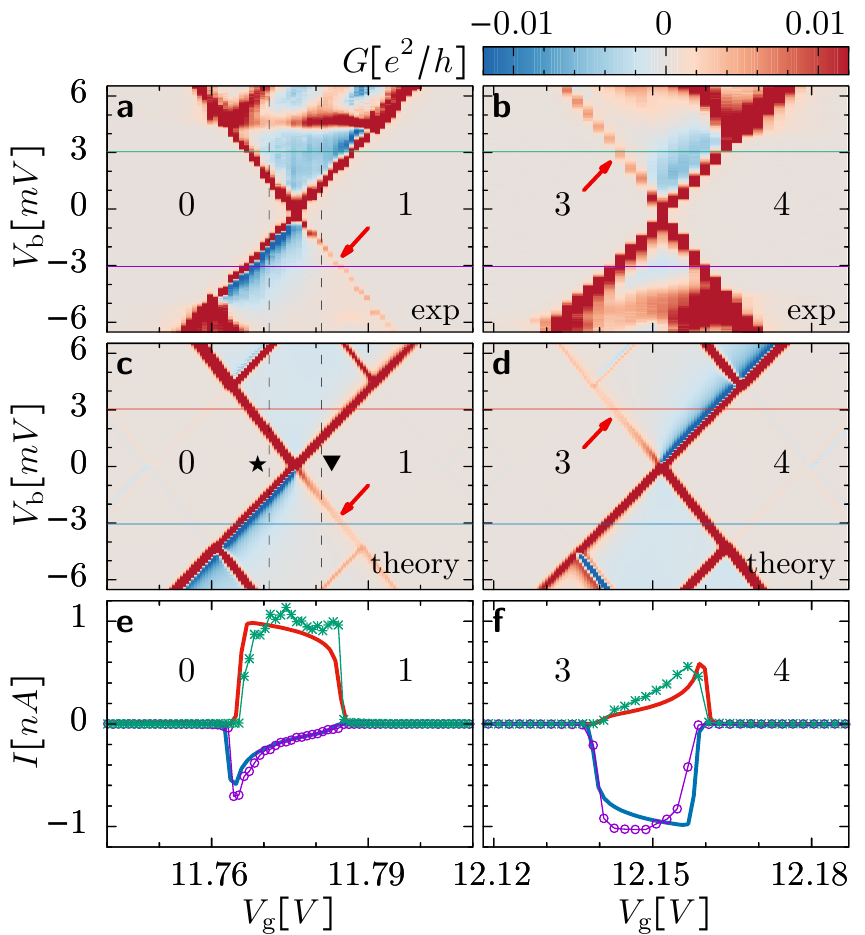}
	\caption[Current suppression and particle-hole symmetry]{
		\bs{a,b}
		Experimental stability diagrams for shell~$1$.  Current suppression features observed for single-electron tunneling in panel (\bs{a}),  also occur for single-hole  tunneling  under reverted bias polarity and mirroring of the gate voltage, as shown in panel (\bs{b}).
		\bs{c,d} Theoretical stability diagrams for the $0 \leftrightarrow 1$ and  $3 \leftrightarrow 4$ dynamical regimes reproducing the experimental observation.
		\bs{e,f}
		Comparison of experimental and numerical current-gate traces at bias voltage set to $V_\mathrm{b}=\pm 3.045$\,mV.
	}
	\label{fig2}
\end{figure}

\begin{figure}
	\centering
 	\includegraphics[width=88mm]{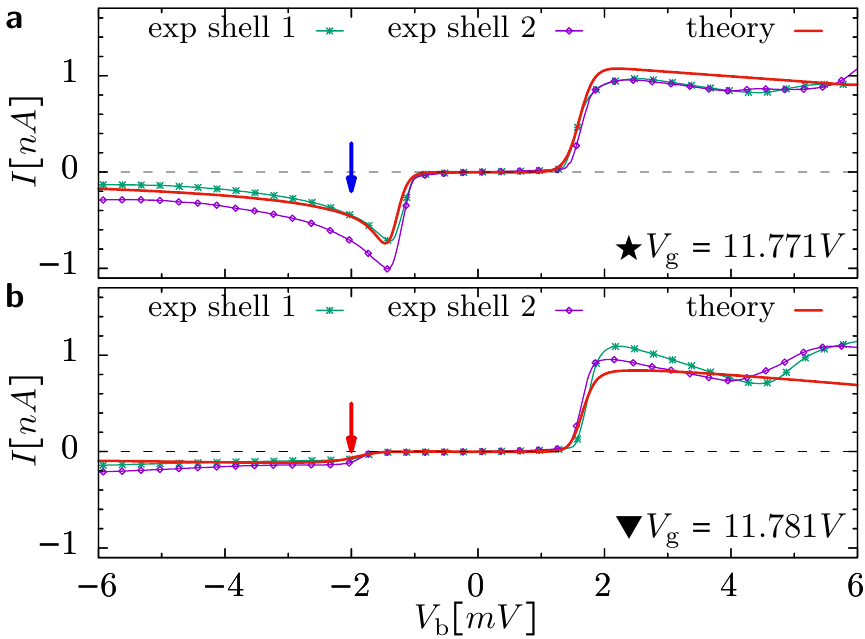}
 	\caption[$I-V$ characteristics in the presence of dark states]{
 		\bs{a,b}
 		Experimental current-bias characteristics around the $0\leftrightarrow1$ resonance for shell~$1$ at voltages  $V_\mathrm{g}=11.771$V (\bs{a}) and  $V_\mathrm{g}=11.781$V (\bs{b}) are  compared to numerical results. These gate voltages correspond to the vertical dashed lines in Fig.~\ref{fig2}. The behavior at positive voltages is similar. At negative bias, however, one observes a pronounced negative differential conductance in panel (\bs{a}) and almost vanishing current in panel (\bs{b}). The more effective coherent population trapping in (\bs{b}) indicates a vanishing precession between DS and CS states, see Fig.~\ref{fig1}a. The current measured in shell 2 displays similar behavior. 
 	}
 	\label{fig3}
\end{figure}

Measurements are performed on a suspended CNT grown on top of prepatterned leads. Such CNTs are usually called ultraclean owing to their low level of impurities \cite{Cao2005}. 
In Fig.~\ref{fig1}(d) we show the experimentally measured differential conductance $G$ of our ultraclean CNT quantum dot as a function of the applied bias voltage $V_\mathrm{b}$ and of a back-gate voltage  $V_\mathrm{g}$. Coulomb diamonds are clearly visible, with a characteristic 4-fold periodicity, a signature of the successive filling of CNT shells with four electrons each.
Noticeably,  three almost identical diamonds are followed by a larger one.
The width of a Coulomb diamond is a measure of the  energy required to fill the CNT with an extra electron, which accounts for charging effects and single-particle properties.
For the small diamonds only a charging energy has to be paid, which indicates almost degenerate states within a shell (negligibly small spin-orbit coupling and valley mixing), as well as a small exchange energy for the middle diamond. For the larger diamonds the shell is full ($N=0$ mod 4), such that filling the CNT with an extra electron requires to pay a charging energy and the mean inter-shell spacing.  
By closer inspection of the current voltage characteristics in shell 1 and shell 2, we observe signatures of  current suppression in the form of faint Coulomb diamond edges and negative differential conductance for the $0\leftrightarrow1$ transition, as indicated by the red and blue arrows, respectively. The same pattern occurs also at the $3\leftrightarrow4$ transition  for opposite bias polarity.
To confirm that such current suppression is due to the formation of a DS, we have compared experimental gate traces and  bias traces with theoretical calculations for the 
 stationary current of a CNT.
  Fig.~\ref{fig1}e shows experimental gate traces for $V_\mathrm{b} = \pm 3.045$\,mV (green/purple line in Fig.~\ref{fig1}d); the numerical calculations are depicted in Fig.~\ref{fig1}f.
The parameters used can be found in Tab.~II of the Methods. The same outcome is seen in the theoretical traces. For further inspection, we have focused on the $0\leftrightarrow1$ and $3\leftrightarrow4$ transitions of shell~$1$. The respective experimental stability diagrams are shown on an enlarged gate voltage scale in Figs.~\ref{fig2}a,b, while the corresponding numerical results are depicted in Figs.~\ref{fig2}c,d. We observe similar behavior as in the experiment within a range of $3$\,mV around zero bias. At larger voltages,  besides  additional diagonal lines  due to excited states,  the experimental curves display  a  horizontal line not understood at present.
Experimental and theoretical gate traces at fixed $V_{\rm b}$ are shown together in Figs.~\ref{fig2}e,f.
At the $0\leftrightarrow1$ resonance both the experimental and theoretical gate traces show a rectangular shaped  current at positive bias, typical of quantum dot behavior in the sequential tunneling regime; at negative bias, however, the current first increases and then gradually decreases as the gate increases, indicating trapping of a single electron. At the $3\leftrightarrow4$ resonance similar current shapes are  observed for opposite bias voltage polarity and upon gate voltage mirroring, a signature of  trapping of a single hole. 
The  dependence of the current on the bias voltage is analyzed in more detail in Fig.~\ref{fig3}. We show the current for the $0 \leftrightarrow 1$ transition at $V_\mathrm{g}=11.771$\,V and $V_\mathrm{g}=11.781$\,V in Figs.~\ref{fig3}a,b, respectively; these positions are marked in Fig.~\ref{fig2} by vertical lines and the corresponding symbols. 
The negative differential conductance and the faint (almost missing) resonant line are highlighted by a blue and red arrow, respectively.
Again, the agreement between theory and experiment is remarkable.  As discussed below, all the characteristic features observed in Figs. \ref{fig1}-\ref{fig3} can be explained in terms of coherent population trapping in a DS, combined with a precessional motion which transfers population between the dark and the coupled state, as sketched in Fig.~\ref{fig1}(a).  
 
\section*{Orbital degeneracy and CNT spectrum} 
The general ingredients to describe charge transport across a quantum dot in the sequential tunneling regime are a tunneling Hamiltonian  coupling the dot to the  electrodes,  and the electronic spectrum of the isolated system in the energy range set by the electrochemical potentials of the lead electrodes. The  presence of orbital degeneracies (or quasi-degeneracies) is decisive for the occurrence of CPT. 
As discussed above, the single particle energy spectrum of a CNT of finite length is fully characterized 
by a shell quantum number $m$ and the pair $(\sigma, \ell_z)$, accounting for the spin and orbital degrees of freedom.
 Curvature-induced spin-orbit coupling and valley-mixing remove the intra-shell degeneracy. The  amplitude of the spin-orbit coupling is largest near the Dirac point and of the order of a fraction of meV. However, it strongly decreases for states away from the bottom of the CNT conduction band \cite{Jespersen2011}, which is the case for the gate voltage range  in which CPT is seen  in our experiment. Similarly, valley mixing due to disorder is strongly suppressed in ultraclean CNTs, and is forbidden by symmetry in CNTs of the zig-zag class \cite{Marganska2015}, which suggests that we have measured such kind of tube in our experiment. We neglect both perturbations in the following. The Hamiltonian we used to describe the CNT quantum dot is thus given in Eq.~(\ref{eq:H}) of the Methods. It consists of three shells $m =0,1,2$, and the only relevant parameters are the inter-shell spacing $\varepsilon_0$, the interaction $U$ accounting for charging effects and the exchange interaction $J$. 
 A many-body state with $N$ particles is thus characterized by its total energy $E$, the total angular momentum $L_z$, and the total spin quantum numbers $S$ and $S_z$, i.e., it has the form  $\vert N, E; S, S_z, L_z \rangle$. In our three-shells model, we have fixed the energy $E_0$ and the particle number  of the configuration with the shell $m=0$ completely full, and the upper two shells $m = 1,2$ completely empty. The $N = 0$ groundstate $\vert 0, E_0; 0, 0, 0 \rangle \equiv \vert 0 \rangle$ is depicted in Tab.~I. 
The $N=1$ groundstate is four-fold degenerate. A basis is the quadruplet of states $\{ \vert 1, E_1; \frac{1}{2}, \sigma, \ell_z \rangle \}$ obtained by adding one electron with quantum numbers $(\sigma,\ell_z)$ on shell $m=1$. These states are also graphically shown in Tab.~I, where we used for them the short-cut notation $\vert \sigma, \ell_z \rangle$. 
Further examples of many-body states  with $N=1,2$ electrons are in the Supplementary Note~I. 
Here we  exemplary focus on CPT at the $0 \leftrightarrow 1$ resonance, which involves the $N=0$ and the $N=1$ groundstates.

\begin{table}[b]
	\label{tab1}
 	\centering
 	\begin{tabular}{l | c c c c c}
 		\hline\hline
 		$m$ & $\vert 0 \rangle$ & $\vert \uparrow,\ell \rangle$ & $\vert \downarrow, \ell \rangle$ & $\vert \uparrow, -\ell \rangle $ & $\vert \downarrow, -\ell \rangle $ \\
 		\hline
 		$\begin{matrix}	2 \\ 1 \\ 0 \end{matrix} $ & $\sx{2}{2}{0}{0}{0}{0}$ & $\sx{2}{2}{u}{0}{0}{0}$ & $\sx{2}{2}{d}{0}{0}{0}$ & $\sx{2}{2}{0}{u}{0}{0}$ & $\sx{2}{2}{0}{d}{0}{0}$ \\
 		\hline\hline
 	\end{tabular}
	\caption[Manybody groundstates]{Configuration corresponding to the $N=0$ groundstate and to the four $N=1$ groundstates in shell $m=1$. }
\end{table}

\section*{Tunneling matrix}
 The tunneling Hamiltonian allows for transitions between dot states with different particle number, and  corresponds to the resonant laser fields inducing transitions in the atomic Lambda systems. Its form, given in Eq.~(\ref{eq:H_tun}) of the Methods, is rather standard.
The associated complex tunneling amplitude $t_{\alpha \vec{k} m \ell_z}$ accounts for the overlap between an electron  wave function in lead $\alpha$, characterized by the momentum ${\vec k}$, and a CNT wavefunction for shell $m$ and angular momentum $\ell_z$ in the contact region. 
We assume that, due to the nanotube curvature, tunneling is local and occurs only through those CNTs atoms closest to the leads. In this case, as discussed in the Methods, the rate matrix $(\boldsymbol{\Gamma}_\alpha^m)_{\ell_z \ell_z'}(\Delta E) := \sum_{\vec{k}}t^*_{\alpha \vec{k} m \ell_z}t_{\alpha \vec{k} m \ell_z'}\delta (\varepsilon_{\vec{k}} -\Delta E)$ is in general non diagonal in the angular momentum basis. 
For a single atom contact, or in the more general surface $\Gamma$-point approximation discussed in the Supplementary Note~III, it takes the simple form $(\boldsymbol{\Gamma}_\alpha^m)_{\ell_z \ell_z'} = \Gamma_\alpha^m (\mathcal{R}_\alpha^m)_{\ell_z \ell_z'} = \Gamma_\alpha^m e^{i \phi_\alpha^m(\ell_z-\ell_z')}$, where the phase $\phi_\alpha^m$ describes a global property of contact $\alpha$ for shell $m$.

\section*{Dynamics and DS trapping}
The dynamical quantity of interest here is the stationary current $I$.  
It follows from the stationary reduced density matrix $\rho^\infty =\lim_{t\to\infty} \rho$ and the current operator $\hat I_\alpha$ at lead $\alpha$  according to  $I:=I_\mathrm{L } = \mathrm{Tr}_\mathrm{CNT} \{\hat{I}_\mathrm{L} \rho^\infty \}=-I_\mathrm{R}$. Notice that this requires to take the trace over the full spectrum of the isolated CNT. 
For weak tunneling coupling, the dynamical equations for $\rho$ are easily obtained from the  Liouville-von Neumann equation for the total density operator by treating the tunneling Hamiltonian as a perturbation \cite{Begemann2008,Darau2009}.
For general values of the gate and bias voltages such equations have to be solved numerically. 
Analytical solutions are possible when the system is tuned near a transition involving only  $N$ and $N+1$ particles groundstates, which is the case of interest here.

Let us consider  the $0 \leftrightarrow 1 $  transition. 
In the  one-dimensional $N=0$ subspace   the density matrix is a number, $\rho_0$.
In the four-dimensional $N=1$ subspace, it is block-diagonal in spin (since spin is conserved during tunneling) but {\it not} in angular momentum. 
The contributions from different spin configurations can be summed up in the dynamical equations yielding a set of coupled equations for $\rho_0$ and a $2\times 2$  matrix $\rho_{1}(E_1)$ \cite{Niklas2017}. Away from the exact resonance (i.e. from the border of the Coulomb diamond), one finds for positive electrochemical potential drop $eV_\mathrm{b} \gg k_\mathrm{B}T$,
\begin{align}
\label{rho}
	\dot{\rho}_1 
	&=
	-\frac{i}{\hbar} \Big[ \hat{H}_\mathrm{LS} , \rho_1 \Big]
	+ 2 \Gamma_\mathrm{L} \mathcal{R}_\mathrm{L} \rho_0
	- \frac{\Gamma_\mathrm{R}}{2} \Big\{ \mathcal{R}_\mathrm{R} , \rho_1 \Big\}
	, \nonumber \\
	\dot{\rho}_0	
	&=
	\Gamma_\mathrm{R} \mathrm{Tr}\left(\mathcal{R}_\mathrm{R} \rho_1\right)
	- 4 \Gamma_\mathrm{L} \rho_0
	,
\end{align}
with $\Gamma_\alpha^m=\Gamma_\alpha$ and the coherence matrices $\mathcal{R}_{\alpha}$. Eq.~(\ref{rho}) describes relaxation governed by the terms $\Gamma_\alpha \mathcal{R}_\alpha$ as well as precession through the Lamb Shift contribution  $\hat{H}_\mathrm{LS}= \frac{\hbar}{2}\sum_\alpha \omega_\alpha \mathcal{R}_\alpha $. The latter originates from virtual processes from the system to the leads \cite{Braun2004,Schultz2009,Donarini2009} and its effect will be discussed later.  
Due to the non diagonal form of the $\mathcal{R}_{\alpha}$ in the angular momentum basis, also the stationary density matrix $\rho_1^\infty$ is not diagonal there. Its diagonalization yields the stationary eigenstates, and the associated eigenvalues define the {\it occupation probabilities} of the eigenstates. 
To proceed, we assume for the coherence matrices the simple form $(\mathcal{R}_\alpha)_{\ell_z \ell_z'}=  e^{i\phi_\alpha (\ell_z-\ell_z')} $ and  $\phi_\mathrm{R}\neq\phi_\mathrm{L}$.
In this case, eigenstates
 of $\rho_1^\infty$ are a decoupled state (DS) at the right  lead  and its associated orthogonal coupled state (CS),  with eigenvalues $1$ and $0$,   respectively. 
Hence, Eq.~(\ref{rho}) predicts CPT with probability one  in the DS, and consequently a vanishing stationary current, see Fig.~\ref{fig1}a.
Accounting also for the spin degree of freedom, the dark/coupled states at the right lead are the linear combinations
\begin{equation}
\label{eq:DS1}
	\left\{ \begin{matrix}
		\vert \mathrm{DS},\sigma;\mathrm{R} \rangle \\
		\vert \mathrm{CS},\sigma; \mathrm{R} \rangle
	\end{matrix}\right.	
	\equiv 
	\frac{1}{\sqrt{2}}\left(
		e^{i \ell \phi_\mathrm{R}}\vert \sigma, \ell \rangle
		\mp
		e^{-i \ell \phi_\mathrm{R}}\vert \sigma, -\ell \rangle
	\right)
	.
\end{equation} 
Notice the explicit dependence on the tunneling phase acquired upon tunneling at the lead R. If $\phi_\mathrm{R}=\phi_\mathrm{L}$, the DS state in Eq.~(\ref{eq:DS1}) is also decoupled at lead L; transport solely occurs through the CS and CPT cannot occur. Thus coherent population trapping requires $\phi_\mathrm{R}\neq \phi_\mathrm{L}$.
 
\begin{figure}
 	\centering
 	\includegraphics[width=88mm]{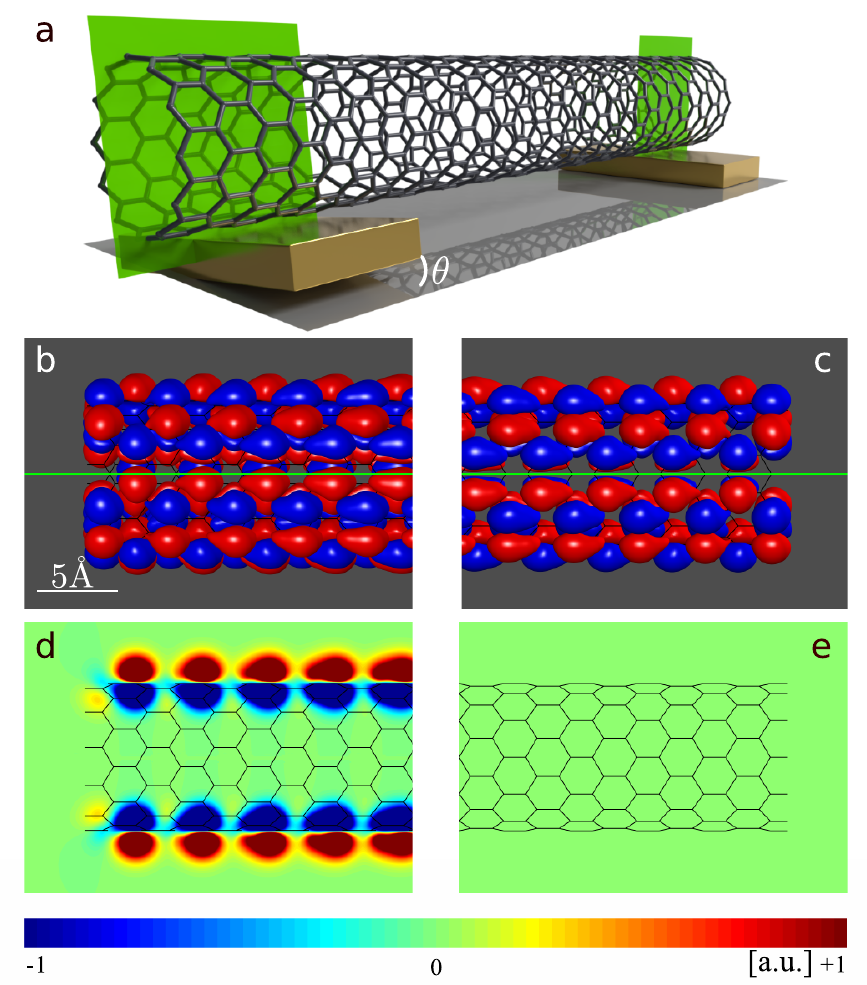}
 	\caption[Dark state of a (12,0) carbon nanotube]{
 		\bs{a}
 		Sketch of the nanotube-lead configuration at the left and right  contacts. The left lead is rotated by an angle $\theta$ with respect to the right lead. The intersection of the green rectangles with the leads define in both leads the contact region.
 		\bs{b,c}
 		Equiamplitude surfaces of the CNT wavefunction as seen from the left (\bs{b}) and right (\bs{c}) contact region. At the left lead, a nodal line is seen which coincides with the green contact line. At the right lead, in contrast, the nodal line and the contact line do not coincide. 
 		\bs{d,e}
 		Projection of the wave function amplitudes on the intersection rectangle. The DS has a vanishing amplitude at the right lead but not at the left lead.
 	}
 	\label{fig4}
\end{figure}

The DS wave function in Eq.~(\ref{eq:DS1}) is explicitly shown in Fig.~\ref{fig4} on the example of  a (12,0) CNT. We have assumed the angular coordinate of the contact atoms at the right and left lead to be rotated by a small  angle $\theta =\pi/24$.     
The DS has a node at the contact positions at the right lead but not at the left lead. The corresponding CS is shown in the Supplementary Figure S1 and has finite weight at both contacts.

Let us now turn to the impact of the Lamb shift term in Eq.~(\ref{rho}). 
It introduces a precession of the Bloch vector in the CS/DS basis of Eq.~(\ref{eq:DS1}) with population transfer between dark and coupled states. The frequencies $\omega_\mathrm{L}$, $\omega_\mathrm{R}$ are given in Eq.~(\ref{eq:omega}) of the Methods. For the situation indicated in Fig.~\ref{fig1}c, $\omega_\mathrm{L}\neq 0$ allows the electrons in a DS to precess into the coupled state and from there to escape, yielding a small stationary current $I=4e\Gamma_\mathrm{L}\rho_0^\infty$.
We find from Eq.~(\ref{rho}) the expression  
\begin{equation}
\label{eq:I_analytic}
	I
	\!=\!
	\frac{
		4e\Gamma_\mathrm{R}\omega_\mathrm{L}^2\cos^2\Delta\phi
	}{	
		8\Gamma_\mathrm{R}^2
		+
		2(\omega_\mathrm{L}-\omega_\mathrm{R})^2
		+
		\omega_\mathrm{L}
		\left(
		 \omega_\mathrm{L}\Gamma_\mathrm{R}/\Gamma_\mathrm{L}
		+
		4 \omega_\mathrm{R}
		\right)
		\cos^2\Delta\phi 
	}
	,
\end{equation}
for $\Delta \phi = (\phi_\mathrm{L} - \phi_\mathrm{R})/2 \neq 0$.
Because the precession frequencies strongly depend on gate and bias voltage, also the effectiveness of CPT does, as clearly seen from the gradual variation of the current in Fig.~\ref{fig3}. 
Here, the traces in panels (a) and (b) correspond to two distinct values of $V_\mathrm{g}$. 
Clearly, at negative bias voltages (corresponding to positive  $eV_\mathrm{b}$), CPT is more pronounced in panel (b) than in panel (a), indicating a smaller $\omega_{\mathrm{L}}$. As the bias polarity is changed, also the role of the precession frequencies is exchanged. The fact that  
 at positive bias the current shows standard Coulomb steps, is because of large $\omega_\mathrm{R}$ (compared to $\omega_\mathrm{L}$) for both of the chosen $V_\mathrm{g}$ values. The dependence of the frequencies $\omega_{\mathrm{L}/\mathrm{R}}$ on $V_\mathrm{b}$ for the same parameters used in the simulation of Fig.~\ref{fig3}a is explicitly shown in the Supplementary Figure~S2. In the Supplementary Figure~S3, we additionally show the dependence of the current when additional inelastic relaxation processes are added in the numerical simulations. 
   
We observe that, due to the particle-hole symmetry of the spectrum with respect to half-filling (shell with $N=2$), the equations for the reduced density operator near the $3 \leftrightarrow 4$ transition immediately follow from Eq.~(\ref{rho}) upon exchanging R with L and  $V_\mathrm{b}$ with  $ - V_\mathrm{b}$. This implies that the CPT features at the $3 \leftrightarrow 4$ transition can be obtained from the ones at the $0 \leftrightarrow 1$ resonance by mirroring of both $V_\mathrm{b}$ and $V_\mathrm{g}$. This is clearly observed in Fig.~\ref{fig2}. 
 
Similar to the $N=0$ groundstate, also the $N=2$ groundstate is a singlet for positive exchange interaction $J$ (see Eq.~(S7) of the Supplementary Note I). However, CPT is  not seen at the $1 \leftrightarrow 2$ and $2 \leftrightarrow 3$ resonances. Whether or not interference occurs depends on the strength of the exchange coupling, since $J/2$ is the separation to the first excited doublet of states which do not form dark states.
For $J \simeq \hbar\Gamma \simeq eV_\mathrm{b}$ the excited doublet is soon into the transport window and no CPT is seen. The $N=2$ states and a special realization of a DS for the case $J=0$ are further discussed in the Supplementary Note I.

\section*{Discussion}
The results presented so far show a remarkable quantitative agreement between the  experimental data and the theoretical predictions, strongly supporting the claim that the observed current suppression features are due to CPT. 
A natural question is how robust CPT is, and under which conditions can it be observed in other CNT-based quantum dots.   
According to our model, the effect is quite generic, as the main requirement are
i)
the presence of a symmetry ${\cal S}$ of the system yielding degenerate energy states (for weak symmetry breaking the level splitting should be smaller than the tunneling broadening  $\Gamma =\Gamma_\mathrm{L} +\Gamma_\mathrm{R}$),
ii)
tunneling matrices being not diagonal in the basis associated to the symmetry ${\cal S}$ and with modulus of the off-diagonal elements of the coherence matrices $\mathcal{R}_\alpha$ close to one,
iii)
strong Coulomb interaction enforcing single electron tunneling. 
 The above requirements are simultaneously met for $n$-doped (electron conduction) CNTs of the zig-zag class, whose bound states have angular momentum (valley) degeneracy. For this nanotube class, symmetry breaking perturbations are spin-orbit coupling and valley mixing. While the former is an intrinsic property of the CNT and decreases away from the band gap, valley  mixing  is due to disorder or to perturbations which break the rotational symmetry. Suspended ultraclean CNTs \cite{Cao2005} show very weak disorder and symmetry breaking mostly occurs due to the presence of contact leads in the non-suspended portion of the tube. In a realistic CNT-device the curvature of the nanotube and some roughness of the contacts causes tunneling to occur locally through few single carbon atoms. This ensures on the one hand that the tunneling matrix is not diagonal in the angular momentum basis, and on the other that the tunneling is a small perturbation and hence that valley mixing is small. Furthermore, weak tunneling makes it easier to reach the sequential tunneling regime, which is typically observed for CNTs in the electron conduction regime \cite{Laird2015}.
 We notice that requirement ii) rules out the possibility that the excited states blocking  recently reported in   a CNT-based quantum dot with broken four-fold degeneracy is due to the CPT discussed here \cite{Harabula2018}. 
   
 We also comment on  interference phenomena observed in the linear conductance of molecular junctions in the strong coupling regime \cite{Guedon2012, Vazquez2012}. 
Since energy conservation is required only within an accuracy set by $\Delta E \approx \hbar\Gamma$, in the strong coupling regime, where $\Gamma$ is the largest scale in the problem, the CPT  based on linear superpositions of quasi-degenerate states of the isolated system looses its significance. Rather,  interference based on a coherent superposition of different  trajectories along the molecule becomes possible \cite{Lambert2015}.
 Similarly, the zero-bias interference discussed for off-resonant transport through single molecules requires energy-nonconserving virtual transitions \cite{Pedersen2014}, and is thus distinct from the interference blockade phenomena presented in this work.  
 
In conclusion, we have demonstrated experimentally and theoretically that CPT can  be realized in artificial atoms by all-electrical means.  In the case considered here, the artificial atom is a CNT quantum with orbital degeneracies resulting from the interplay of the tubular nanotube geometry and the underlying graphene honeycomb lattice. However, the phenomenon is rather generic and is expected to occur in other highly symmetric quantum dot systems, e.g. in organic molecules in STM setups \cite{Donarini2012} or in symmetric triangular quantum dots \cite{Niklas2017} in the weak tunneling regime. Trapping in a DS can be achieved by simple tuning of the bias or gate voltage.


\section*{Methods}
\subsection*{Sample details}
On a highly p-doped Si substrate with a $500$\,nm thermally grown SiO$_{2}$ layer, ring-like electrode structures were defined by electron beam lithography and evaporation of $20$\,nm Re and $40$\,nm Co. A catalyst for the CVD process was deposited in the center of the electrode ring structure to increase the chance of a CNT connecting the contacts. The CNT growth was performed as the last sample fabrication step to ensure the production of ultra clean devices \citep{Cao2005}. Neither magnetoresistance nor magneto-optical Kerr effect 
(MOKE) measurements showed magnetic behavior of the Re/Co contacts after CVD, ruling out spin valve effects with a similar $I-V$ characteristics \citep{Braun2004}. The current as well as the differential conductance were measured using an Ithaco DL1211 IV-converter.
\subsection*{Model}
The single-electron transistor setup is modeled by the total Hamiltonian $\hat{H}=\hat{H}_\mathrm{CNT} + \hat{H}_\mathrm{leads} + \hat{H}_\mathrm{tun}$, describing the CNT quantum dot (QD) weakly coupled to leads by a tunneling Hamiltonian $\hat{H}_\mathrm{tun}$.

\sect{Isolated carbon nanotube}
The CNT-QD consists of a set of longitudinal modes, called shells, with shell index $m$. The electrons have both spin $\sigma \in \{\uparrow, \downarrow\}$ and valley $\ell_z \in \{\ell,-\ell\}$ degrees of freedom and, since spin-orbit coupling and valley mixing are negligible  in the experiment, the single-particle spectrum of a single shell is assumed to be $4$-fold degenerate. We include a charging term $U$ and  the exchange interaction $J$. The latter strongly depends on the CNT chirality and radius  and was assumed to be  $10\mu$eV. The CNT-QD Hamiltonian reads \citep{Secchi2009, Laird2015}
\begin{align}
\label{eq:H}
	\hat{H}_\mathrm{CNT}
	=&
	\sum\limits_{m \ell_z}
	\left(m\varepsilon_0 - e\alpha_\mathrm{g}V_\mathrm{g}\right)
	\hat{n}_{m \ell_z}
	+
	\frac{U}{2}\hat{N}^2
	\nonumber \\
	&+ J\sum\limits_m\left(\hat{\mathbf{S}}_{m\ell} \cdot \hat{\mathbf{S}}_{m-\ell} + \frac{1}{4}\hat{n}_{m\ell}\hat{n}_{m-\ell}\right)
	,	
\end{align}
where in the numerical calculations only 3 shells ($m\in\{0,1,2\}$) are considered, see Tab.~I.
The gate voltage $V_\mathrm{g}$ applied with a level arm $\alpha_\mathrm{g}$ ensures particle-hole symmetry of Eq.~(\ref{eq:H}) with respect to shell $m = 1$ for $e\alpha_\mathrm{g}V_\mathrm{g} = \varepsilon_0 + 6U + J/4$.
Further, the occupation operator $\hat{n}_{m \ell_z} = \sum_\sigma d_{m \ell_z\sigma}^\dagger d_{m \ell_z\sigma }$, and the spin operator $\hat{\mathbf{S}}_{m \ell_z} = \frac{1}{2}\sum_{\sigma \sigma^\prime}d^\dagger_{m \ell_z \sigma}\boldsymbol{\sigma}_{\sigma \sigma^\prime}d_{m \ell_z \sigma^\prime}$ are defined in terms of creation (destruction) operators $d^{(\dagger)}_{m\ell_z\sigma}$ of an electron in shell $m$ with angular momentum $\ell_z$ and spin projection $\sigma$.

The CNT Hamiltonian can be diagonalized analytically by using the  basis corresponding to the eigenststates of the total particle number $\hat{N} = \sum_{m \ell_z}\hat{n}_{m \ell_z}$, total spin $S^2 = \sum_{m \ell_z} \hat{\mathbf{S}}_{m \ell_z}^2$, total spin projection $S_z = \frac{1}{2}\sum_{m \ell_z \sigma}\sigma d_{m \ell_z\sigma}^\dagger d_{m \ell_z\sigma }$, and total angular momentum operator $L_z = \sum_{m \ell_z}\ell_z  \hat{n}_{m \ell_z}$. Accordingly, many-body states are uniquely defined by the vector set  $\{\vert N ,E; S, S_z, L_z \rangle\}$. In our three-shell model, the $N=0$ groundstate corresponds to the shell $m=0$ being completely full, and the $N=1$ groundstate is given by the quadruplet $\{\vert 1 ,\varepsilon_0; \frac{1}{2}, \sigma, \ell_z \rangle \}$, as shown in Tab.~I. 
Examples of many-body states with occupation $N=2$ can be found in the Supplementary Note I.

\sect{Leads Hamiltonian}
The electrons in the leads are considered as fermionic reservoirs of non-interacting electrons at temperature $T$ 
and chemical potentials $\mu_\mathrm{L}=\mu_0 +e \eta V_\mathrm{b}$, $\mu_\mathrm{R}=\mu_0 + e(\eta-1)V_\mathrm{b}$ for the left (L) and right (R) lead, respectively, with $V_\mathrm{b}$ the applied bias voltage. The parameter  $0 < \eta < 1$ accounts for an asymmetric bias drop at the two leads. For the investigated setup a good fit to the data was obtained for almost symmetric potential drop ($\eta =0.55$). For simplicity, the  underlying Hamiltonian is chosen to be the one of a non-interacting electron gas
$H_\mathrm{leads} = \sum_{\alpha\vec{k}\sigma} \varepsilon_{\vec{k}} c^\dagger_{\alpha\vec{k}\sigma}c_{\alpha\vec{k}\sigma}$, where $c^{(\dagger)}_{\alpha\vec{k}\sigma}$ destroys (creates) an electron in lead $\alpha\in\{\mathrm{L},\mathrm{R}\}$ with momentum $\vec{k}$ and spin projection $\sigma$.

\sect{Tunneling Hamiltonian}
The leads are weakly coupled to the CNT via the tunneling Hamiltonian
\begin{equation}
\label{eq:H_tun}
	H_\mathrm{tun}
	=
	\sum\limits_{\alpha \vec{k} m \ell_z \sigma}
	t_{\alpha \vec{k} m \ell_z}
	d^\dagger_{m \ell_z \sigma}
	c_{\alpha \vec{k} \sigma}
	+ \mathrm{h.c.}
	.
\end{equation}
Thus, $H_\mathrm{tun} = \sum_{\alpha}H_{\mathrm{tun},\alpha}$ removes (adds) an electron from the left/right lead ($\alpha = \mathrm{L}/\mathrm{R}$) with momentum $\vec{k}$, energy $\varepsilon_{\vec{k}}$  and spin $\sigma$ and it adds (removes) an electron in the dot with the same spin $\sigma$ in the state $(m, \ell_z, \sigma)$. 

The tunneling Hamiltonian Eq.~(\ref{eq:H_tun}) rules the dynamics of the coupled CNT-leads system. For weak coupling, it can be treated as a perturbation. Expanding the Liouville von Neumann equation for the total density operator up to second order in $H_\mathrm{tun}$, and taking the trace over the reservoirs, the equations for the reduced density operator are obtained \cite{Begemann2008, Darau2009}. They are ruled by a tunneling kernel, which, according to Eq.~(\ref{eq:H_tun}), has the form 
\begin{equation}
	(\boldsymbol{\Gamma}_\alpha^m)_{\ell_z \ell_z'}(\Delta E)
	:=
	\frac{2\pi}{\hbar}
	\sum\limits_{\vec{k}}
	t^*_{\alpha \vec{k} m \ell_z}
	t_{\alpha \vec{k} m \ell_z'}
	\delta (\varepsilon_{\vec{k}} -\Delta E)
	,
\label{eq:tun_matr}
\end{equation}
and is in general non diagonal in the angular momentum basis. Rather $(\boldsymbol{\Gamma}_\alpha^m)_{\ell_z \ell_z'} = \Gamma_\alpha^m (\mathcal{R}_{\alpha})_{\ell_z\ell_z^\prime}$, where the coherence matrices 
$\mathcal{R}_{\alpha}$ are 
  hermitian. Explicitly,  $(\mathcal{R}_{\alpha})_{\ell_z\ell_z} = 1$ while the off-diagonal elements satisfy $0 \le \vert (\mathcal{R}_{\alpha})_{\ell_z,-\ell_z}\vert \le 1 $.
As discussed in the Supplementary Note III, the $\mathcal{R}_{\alpha}$ become exactly diagonal in the limit in which all the atoms of the CNT are equally coupled to the leads.
In the opposite case of a single atom contact along the CNT circumference, or in the more general so-called surface $\Gamma$-point approximation, see Supplementary note III, one finds the optimal coherence, and the coherence matrices at lead $\alpha = {\rm L,R}$ assume the form $(\mathcal{R}_{\alpha})_{\ell_z\ell_z^\prime} = e^{i\phi_\alpha (\ell_z-\ell_z')}$ used in the main manuscript.
Then, in the dark and coupled states basis of Eq. (\ref{eq:DS1}),  $	\mathcal{R}_{\rm R}$ becomes diagonal but not $	\mathcal{R}_{\rm L }$. Explicitly, 
\begin{align}
	\mathcal{R}_\mathrm{R}
	&=
	\begin{pmatrix}
		0 & 0 \\
		0 & 2
	\end{pmatrix}
	, \nonumber \\	
	\mathcal{R}_\mathrm{L}
	&=
	2
	\begin{pmatrix}
		\sin^2 \Delta\phi & -i\sin \Delta\phi \cos \Delta\phi \\
		i\sin \Delta\phi \cos \Delta\phi & \cos^2 \Delta\phi
	\end{pmatrix}
	.
\end{align}

The good agreement found between theory and experiment suggests that in our set-up the coherence matrices have off-diagonal elements with modulus close to one in the angular momentum basis, i.e. tunneling occurs only at few atomic position where the CNT is closest to the leads.  

\subsection*[Tunneling dynamics at the 0-1 resonance]{Tunneling dynamics at the $0\leftrightarrow1$ resonance}
For the parameters used in this work, the single particle groundstate is four-fold degenerate, while the two particles groundstate is only at a distance $J/2 \approx \hbar\Gamma $ from the doublet of first excited states, as discussed in the Supplementary Note I. Thus, the secular approximation is non-valid when transitions involving the low-lying two-particles states are considered. To account for the non-secular contributions, the equations for the reduced density matrix have thus been derived along the lines discussed in \cite{Darau2009}. 
In addition, a relaxation term was added to account for inelastic processes due to e.g. phonons. Such equations have been then solved numerically within the three-shell approximation for the CNT spectrum discussed above, and by including all excitations up to a cut-off energy $1.5\,\varepsilon_0$.

Here we focus on the dynamics near the $0 \leftrightarrow 1$ resonance, where only the $0$ and $1$ particle groundstates play a role and an analytical treatment is possible. Furthermore, the secular approximation holds true yielding for the stationary reduced operator the equation 
\begin{equation}
   0
   =
   \mathcal{L}\rho^\infty
   =
   -\frac{i}{\hbar}\left[
   \hat{H}_\mathrm{CNT}
   +\hat{H}_\mathrm{LS},\rho^\infty
   \right]
   +\mathcal{L}_\mathrm{tun} \rho^\infty
   +\mathcal{L}_\mathrm{rel} \rho^\infty
   ,
\end{equation}
where $\mathcal{L}$ is the Liouville superoperator which contains the system internal dynamics in a commutator structure together with the Lamb shift contribution $\hat{H}_\mathrm{LS}$, a tunneling part $\mathcal{L}_\mathrm{tun}$ and a relaxation part $\mathcal{L}_\mathrm{rel}$. 
For positive potential drop $eV_\mathrm{b}$, the equation of motion for the $\rho_{0/1}$ subblocks, where the contribution over spin configurations has been summed, reads 
\begin{align}
\label{rho_rel}
	0
	=
	\dot{\rho}_1 
	&=
	-\frac{i}{\hbar} \Big[ \hat{H}_\mathrm{LS} , \rho_1 \Big]
	+
	2 \Gamma_\mathrm{L} \mathcal{R}_\mathrm{L} \rho_0
	-
	\frac{\Gamma_\mathrm{R}}{2} \Big\{ \mathcal{R}_\mathrm{R} , \rho_1 \Big\}
	\nonumber \\
	&\quad 
	-
	\Gamma_\mathrm{rel}[\rho_1 - \rho_{1,\mathrm{th}}\mathrm{Tr}(\rho_1)]
	,
	\nonumber \\
	0
	=
	\dot{\rho}_0	
	&=
	\Gamma_\mathrm{R} \mathrm{Tr}\left(\mathcal{R}_\mathrm{R} \rho_1\right)
	-
	4 \Gamma_\mathrm{L} \rho_0
	,
\end{align}
where
$\rho_{1,\mathrm{th}} = 1/2$ is the  thermal density matrix for the one-electron sub-block. The dynamics is thus fully governed by the coherence matrices $	\mathcal{R}_\alpha$. 
They enter in the tunneling terms $\Gamma_\alpha \mathcal{R}_\alpha$ as well as in the angular Lamb shift Hamiltonian $\hat{H}_\mathrm{LS} = \hbar \sum_\alpha \omega_\alpha \mathcal{R}_\alpha/2$.
The latter introduces a precession with frequencies \cite{Darau2009}
\begin{equation}
\label{eq:omega}
	\omega_\alpha(V_\mathrm{g},V_\mathrm{b})
	=
	\frac{\Gamma_\alpha}{\pi}
	\left[
		p_\alpha(-eV_\mathrm{g})
		-
 		p_\alpha\left(U-\frac{J}{2} -eV_\mathrm{g}\right)
	\right]
	,
\end{equation}
where we chose an offset in the gate voltage such that the resonance is exactly at $V_\mathrm{g}=0$.
Furthermore $p_\alpha\left(\Delta E\right) := -\operatorname{Re} \psi\left[1/2+i(\Delta E - \mu_\alpha)/2\pi k_\mathrm{B}T\right]$ where $\psi$ is the digamma function.
These precession frequencies clearly depend on the gate voltage and, via the chemical potentials, also on the bias voltage. This dependence is further analyzed in the Supplementary Note II. Notice also the dependence of $\omega_\alpha$ on the charging energy $U$ and exchange interaction $J$, that results  from  virtual processes involving also the two electron subspaces. The above equations are general.
To proceed, we assume full coherence, such that $(\mathcal{R}_{\alpha})_{\ell_z\ell_z^\prime} = e^{i \phi_\alpha (\ell_z-\ell_z^\prime)}$. Then, when expressed in matrix form, Eq.~(\ref{rho_rel}) yields the set of coupled  equations for populations and coherences
\begin{align}
	\dot{\rho}_0
	&=
	-4\Gamma_\mathrm{L} \rho_0
	+\Gamma_\mathrm{R} ( \rho_{\ell\ell} + \rho_{-\ell-\ell} )
	\nonumber \\ & \hspace{2cm}
	+\Gamma_\mathrm{R} ( e^{-2i\phi_R} \rho_{\ell-\ell} +  e^{2i\phi_R} \rho^{*}_{\ell-\ell} )
	,
	\nonumber \\
	\dot{\rho}_{\ell\ell}
	&=
	-\Gamma_\mathrm{R} \rho_{\ell\ell}
	+2\Gamma_\mathrm{L} \rho_0
	-\left(
		\frac{\Gamma_\mathrm{R}}{2} e^{-2i\phi_R}
		- i \tilde{\omega}^*
	\right) \rho_{\ell-\ell}
	\nonumber \\ & 
	-\left(
		\frac{\Gamma_\mathrm{R}}{2} e^{2i\phi_R}
		+ i \tilde{\omega}
	\right) (\rho_{\ell-\ell})^*
	- \frac{\Gamma_\mathrm{rel}}{2}\left(
		\rho_{\ell\ell}
		- \rho_{-\ell-\ell}
	\right)
	,
	\nonumber \\
	\dot{\rho}_{-\ell-\ell}
	&=
	-\Gamma_\mathrm{R} \rho_{-\ell-\ell}
	+2\Gamma_\mathrm{L} \rho_0
	-\left(
		\frac{\Gamma_\mathrm{R}}{2} e^{-2i\phi_R}
		- i \tilde{\omega}^*
	\right) \rho_{\ell-\ell}
	\nonumber \\ &
	-\left(
		\frac{\Gamma_\mathrm{R}}{2} e^{2i\phi_R}
		+ i \tilde{\omega}
	\right) (\rho_{\ell-\ell})^*
	- \frac{\Gamma_\mathrm{rel}}{2}\left(
		\rho_{-\ell-\ell}
		- \rho_{\ell\ell}
	\right)
	,
	\nonumber \\
	\dot{\rho}_{\ell-\ell}
	&=
	- (\Gamma_\mathrm{R} + \Gamma_\mathrm{rel})\rho_{\ell-\ell}
	-\left(
		\frac{\Gamma_\mathrm{R}}{2} e^{2i\phi_R}
		- i \tilde{\omega}
	\right) \rho_{\ell\ell}
	\nonumber \\ & \hspace{7mm}
	+2\Gamma_\mathrm{L} e^{2i\phi_R} \rho_0
	-\left(
		\frac{\Gamma_\mathrm{R}}{2} e^{-2i\phi_R}
		+ i \tilde{\omega}^*
	\right) \rho_{-\ell-\ell}
	,
\end{align}
with $\tilde{\omega} = \omega_\mathrm{L} e^{2i\phi_\mathrm{L}} + \omega_\mathrm{R} e^{2i\phi_\mathrm{R}}$.
These equations can be solved in the stationary limit with $\dot{\rho}_0=0$ and $\dot{\rho}_{\ell,\ell'}=0$, together with the condition $\mathrm{tr}\rho = \rho_{\ell\ell}+\rho_{-\ell-\ell}+\rho_0 = 1$.
If $\phi_\mathrm{L}=\phi_\mathrm{R}$ and $\Gamma_\mathrm{rel} = 0$ the dark state is completely decoupled from the dynamics and therefore the stationary solution is not uniquely defined but rather depends on the initial state. Any finite relaxation rate or phase difference solves this problem.
Since the general expression for the current is quite lengthy, we focus on two limiting cases.
For vanishing relaxation rate but $\Delta \phi = (\phi_\mathrm{L} - \phi_\mathrm{R})/2 \neq 0$  we obtain the current $I(\Gamma_\mathrm{rel}=0) = 4e\Gamma_\mathrm{L}\rho^0_\infty$ given in Eq.~(\ref{eq:I_analytic}) of the main manuscript.

As mentioned before, this expression is only valid for a non-zero phase difference.
For negative bias the current is the same upon exchanging L$\leftrightarrow$R and the overall sign.
This expression shows that the current is completely suppressed for $\Delta\phi = \pi/2$ despite finite Lamb shift.
The second limit that can be analyzed is a finite relaxation rate and $\Delta\phi = 0$.
Interestingly, the dependence of the current on the relaxation rate drops completely and we obtain 
\begin{equation}
	I(\Delta\phi=0)
	=
	e\frac{4\Gamma_\mathrm{L}\Gamma_\mathrm{R}}
	{4\Gamma_\mathrm{L}+\Gamma_\mathrm{R}}
	.
\end{equation}
Notice that this same expression for the current holds true for transport through four fold degenerate levels in the absence of interference. The behavior of the current as a function of the phase difference $\Delta \phi$ is further analyzed in the Supplementary Note II. 

\subsection*{Parameters}
All parameters used for numerical calculation can be found in the following Tab.~II.
\begin{table}[h]
	\centering
	\begin{tabular}{c r r r}
		\hline\hline
		parameter 					& shell $0$ 	& shell $1$		& shell $2$ \\
		\hline
		$\varepsilon_0$				& 				& $4.35$meV		& \\
		$U$							& 				& $20$meV		& \\
		$J$							& 				& $10\mu$eV 	& \\
		$k_\mathrm{B}T$				& 				& $0.5$meV		& \\
		$\hbar\Gamma_\mathrm{R}$	& $2 \mu eV$	& $10\mu$eV		& $10\mu$eV \\
		$\hbar\Gamma_\mathrm{L}$	& 				& $4\mu$eV		& \\
		$\hbar\Gamma_\mathrm{rel}$	& 				& $0.1\mu$eV	& \\
		$\Delta\phi$				& $0.01\pi$ 	& $0.11\pi$		& $0.07\pi$ \\
		$\eta$						& 				& $0.55$		& \\
		\hline\hline
	\end{tabular}
	\caption[Parameters]{Numerical parameters to fit all three shells of the experiment. Only $\Gamma_\mathrm{L}$ and $\Delta\phi$ vary with the shell, all other parameters are the same for all shells.}
\end{table}

\section*{Acknowledgements}

The authors acknowledge financial support by the Deutsche Forschungsgemeinschaft via SFB 1277 and the IGK ``Topological Insulators''. The authors thank A. H{\"u}ttel and M. Marganska for fruitful discussions.

\section*{Author contributions}

A.D., M.N and M.G. performed the theoretical analysis, with numerical simulations carried out by M.N..
The experiments were designed and analyzed by N.P. and C.S..
The devices were fabricated by M.S. and N.P., who also conducted the experiment.
The manuscript and Supplementary Material was mainly written by M.G. and M.N. with critical comments provided by all authors.

\section*{Competing financial interests}
The authors declare no competing financial and non-financial interests.


\onecolumngrid
\clearpage
\pagebreak
\setcounter{equation}{0}
\setcounter{figure}{0}
\setcounter{table}{0}
\setcounter{page}{1}
\makeatletter

\renewcommand{\thefigure}{S\arabic{figure}}
\renewcommand\theequation{S\arabic{equation}}
\renewcommand*\vec[1]{\ensuremath{\boldsymbol{#1}}}

\begin{center}
	\large\bs{Supplementary Information}
\end{center}

\section{Eigenstates and dark states of a carbon nanotube}

We discuss many-body configurations of a CNT quantum dot described by the  Hamiltonian
\begin{equation}
	\hat{H}_\mathrm{CNT}
	=
	\sum\limits_{m \ell_z}
	m\varepsilon_0
	\hat{n}_{m \ell_z}
	+
	\frac{U}{2}\hat{N}^2
	+
	J\sum\limits_m \left(
		\hat{\mathbf{S}}_{m\ell} \cdot \hat{\mathbf{S}}_{m-\ell}
		+
		\frac{1}{4}\hat{n}_{m\ell}\hat{n}_{m-\ell}
	\right)
	,	
\end{equation}
also given in Eq. (4) of the main manuscript. Here,  $m=0,1,2$ runs over three longitudinal shells, $\varepsilon_0$ is the shell spacing, $U$ accounts for charging effects, and $J$ is the exchange interaction. The occupation operator $\hat{n}_{m\ell_z}$ defines the number of electrons in shell $m$ with angular momentum $\ell_z$, where a summation over the spin degree of freedom is performed.  The eigenstates $\ket{N,E; S,S_z,L_z}$ of the CNT Hamiltonian are classified according to their occupation $N$ with respect to a reference state $\vert 0\rangle$ given below, their energy $E$ and total spin and angular momentum quantum numbers $S,\,S_z$ and $L_z$. 
The reference state $\ket{0}$ has the shell $m=0$ completely full, and the upper two shells $m = 1,2$ completely empty and corresponds to the $N=0$ groundstate: 
\begin{equation}
	\ket{0, E_0; 0, 0, 0}
	\equiv
	\ket{0}
	=
	\sx{2}{2}{0}{0}{0}{0}
	.
\end{equation}
The left (right) states are for angular momentum values $\ell_z = +(-)\,\ell$ and the up/down arrows indicate opposite spin direction.
We fix the energy of this state to be $E_{0}$. Many-body excited states with $N=0$ are obtained by creating electron-hole pairs starting from this ground state configuration for fixed electron number.  In   Table~\ref{tab:CNT_states_01} we show both the groundstate as well as all the  first excited states, which have energy $\varepsilon_0 - J/2$ above the groundstate. 
The $N=1$ groundstate is trivially four-fold degenerate
\begin{equation}
\label{GSN1}
	\vert 1, E_{1_0} = \epsilon_0; \frac{1}{2}, \pm\frac{1}{2}, \pm\ell \rangle
	\equiv
	\vert \pm\frac{1}{2}, \pm\ell \rangle
	=
	\left\{
		\sx{2}{2}{u}{0}{0}{0}
		,
		\sx{2}{2}{d}{0}{0}{0}
		,
		\sx{2}{2}{0}{u}{0}{0}
		,
		\sx{2}{2}{0}{d}{0}{0}
	\right\}
	,
\end{equation}
with these states playing a crucial role in the CPT effect discussed in this work.
In Tab. \ref{tab:CNT_states_01}, \ref{tab:CNT_states_12} the lowest excited states with  $N=1$, corresponding to the energies $\varepsilon_0-J$,  $\varepsilon_0 -J/2$, and $\varepsilon_0$ are also shown. Finally, in tables   Tabs.~\ref{tab:CNT_states_12}, \ref{tab:CNT_states_3} we depict the lowest energy states with $N=2,3$. Notice that in the numerics we have considered  only states up to a cut-off of $1.5\varepsilon_0$ above the respective groundstate.

From the many-body energies, addition energies $E_N^{\rm add}=\mu(N+1)-\mu(N)=E_{N+1}-2E_N + E_{N-1} $ are easily calculated. 
We find 
$E_0^\mathrm{add} = \varepsilon_0 + U - J/2$, $E_1^\mathrm{add} = U - J/2$, $E_2^\mathrm{add} = U + 3J/2$ and $E_3^\mathrm{add} = U - J/2$ which in turn define the heights of the Coulomb diamonds.
\subsection{Dark states for one electron}

Eq.~(\ref{GSN1}) allows one to construct linear combinations $\vert  \mathrm{DS}, \sigma; \alpha \rangle $ of the single-particle groundstates $\vert \sigma, \ell_z \rangle$  which are decoupled at given positions $\vec{r}_\alpha$, 
and hence may act as dark states (DS) for a given bias polarity. Such states have the generic form 
$\vert  \mathrm{DS}, \sigma; \alpha \rangle = a(\vec{r}_\alpha) \vert \sigma, \ell \rangle + b(\vec{r}_\alpha) \vert \sigma, -\ell \rangle $, where the coefficients satisfy the normalization condition  $\vert a (\vec{r}_\alpha)\vert ^2 +\vert b (\vec{r}_\alpha)\vert^2=1$ and are determined through the requirement  
\begin{equation}
\label{DS_condition}
	\langle 0 \vert d_{\alpha \sigma} \vert \mathrm{DS}, \sigma; \alpha \rangle
	\equiv
	0
	, 
\end{equation}
where $d_{\alpha \sigma}$ destroys a CNT electron of spin $\sigma$ at position $\vec{r}_\alpha$.
 We express such operator in the angular momentum basis, $d_{\alpha \sigma}=\sum_{m \ell_z} \langle \vec{r}_\alpha \vert m \ell_z \sigma \rangle d_{m \ell_z \sigma} $, and observe that the orbital part $\phi_{m \ell_z \sigma}(\vec{r}_\alpha) = \langle \vec{r}_\alpha \vert m \ell_z \sigma \rangle $ of the CNT wave function is complex, $\phi_{m \ell_z \sigma}(\vec{r}_\alpha) = \vert \phi_{m \ell_z \sigma}(\vec{r}_\alpha) \vert e^{i\theta_\alpha (m,\ell_z)}$.  
Furthermore, $\vert \phi_{m \ell_z \sigma}(\vec{r}_\alpha) \vert = \vert \phi_{m - \ell_z -\sigma}(\vec{r}_\alpha) \vert$ due to time-reversal symmetry.
Then, insertion in Eq.~(\ref{DS_condition}) yields for the coefficients the simple form $a ({\bf r}_\alpha)  =e^{-i \theta_\alpha (m,\ell)}/\sqrt{2}$, $b ({\bf r}_\alpha)=-e^{-i \theta_\alpha (m,-\ell)}/\sqrt{2}$. 
Introducing the angles $\bar\theta_\alpha = [\theta_\alpha (m,\ell) + \theta_\alpha(m,-\ell)]/2$, and $\Delta \theta_\alpha = [\theta_\alpha (m,\ell) - \theta_\alpha(m,-\ell)]/2 $, we find - apart from an overall phase - for the DS the form given in Eq.~(2) of the main text, 
\begin{align}
	\label{DS_graphic}
	\vert \mathrm{DS}, \frac{1}{2};\alpha  \rangle
	&=
	\frac{e^{-i\bar\theta_\alpha}}{\sqrt{2}}\Bigg(
		e^{i \ell \phi_\alpha} \sx{2}{2}{u}{0}{0}{0}
		-
		e^{-i \ell \phi_\alpha} \sx{2}{2}{0}{u}{0}{0}
	\Bigg)
	, \nonumber \\
	\vert {\rm DS}, -\frac{1}{2};\alpha \rangle
	&=
	\frac{e^{-i\bar\theta_\alpha}}{\sqrt{2}}\Bigg(
		e^{i \ell \phi_\alpha} \sx{2}{2}{d}{0}{0}{0}
		-
		e^{-i \ell \phi_\alpha} \sx{2}{2}{0}{d}{0}{0}
	\Bigg)
	,
\end{align}
where $\phi_\alpha \equiv -\Delta \theta_\alpha/\ell$.
Given the form in Eq.~(\ref{DS_graphic}) for the DS at lead $\alpha$, one finds  $\langle 0 \vert d_{\bar{\alpha}\sigma} \vert  \mathrm{DS},\sigma;\alpha \rangle \propto \sin(\ell(\phi_\mathrm{R}-\phi_\mathrm{L})) $ for the matrix element involving the destruction operator at the opposite lead $\bar\alpha$. Thus the requirement $\phi_\mathrm{R}\neq\phi_\mathrm{L}$ is necessary for a dark state to have vanishing transition amplitude only at one lead.

\begin{figure}
	\centering
	\includegraphics[width=15cm]{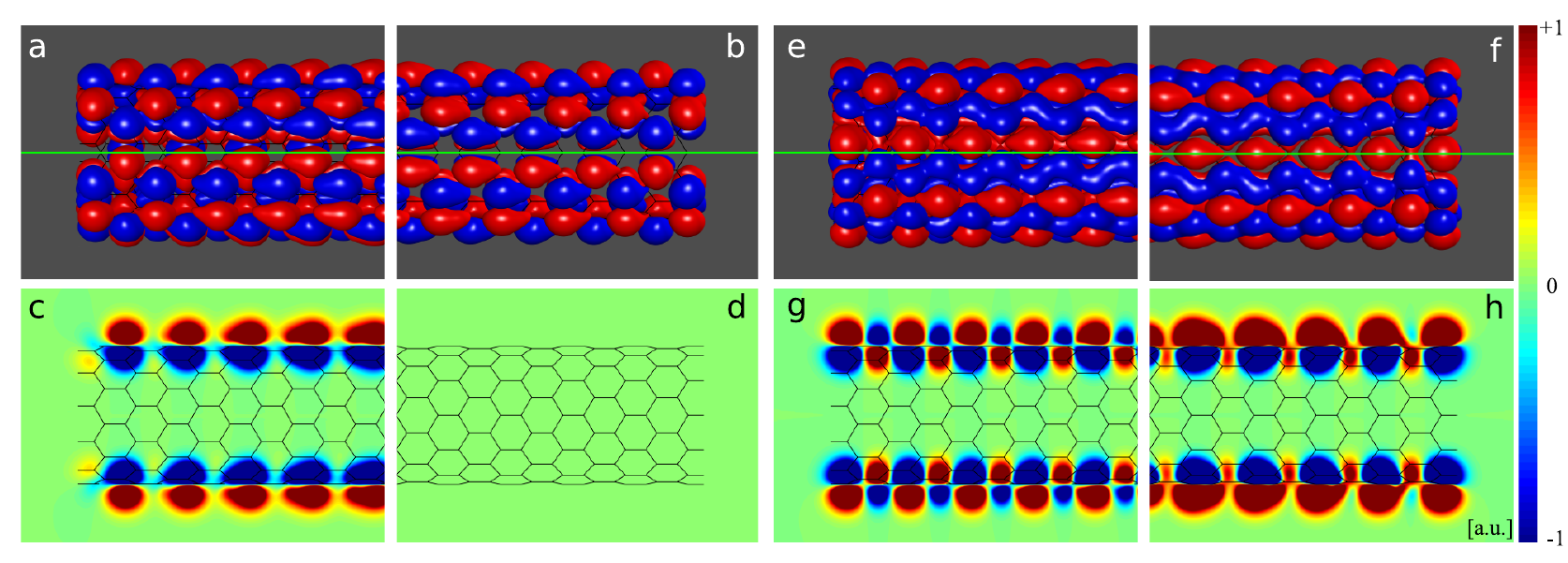}
	\caption[Dark and coupled states of a (12,0) CNT]{
		\bs{a,b}
		Equiamplitude surface of the wave function associated to a decoupled state (DS) near the left (\bs{a}) and right (\bs{b}) ends.
		\bs{c,d}
		Projection of the wave function on the intersection rectangle passing through the CNT and perpendiculat to the lead in the contact region (see Fig.~4a of the main text).
		\bs{e-h} The same features discussed for the DS are here shown for its associated orthogonal coupled state (CS).
	}
	\label{figS1}
\end{figure}

An example of dark and coupled states is shown in Fig.~\ref{figS1} for the case of a $(12,0)$ CNT discussed in the main text (see Figure 4 there). The CNT was chosen to be 100 unit cells long, corresponding to about $50$nm.
Moreover, the states shown correspond to the first excited state above the band gap.

\subsection{Two-electrons dark states}
The results of the $0\leftrightarrow1$ transitions can be easily extended to higher electron numbers since a dark state which blocks transitions to $2$-electrons states can also be constructed. The exchange interaction in the CNT Hamiltonian in Eq.~(3) with antiferromagnetic coupling $J$ can be expressed as
\begin{equation}
	\hat{H}_\mathrm{exch}
	=
	J
	\sum\limits_m
	\left(
		\hat{\mathbf{S}}_{m\ell}
		\cdot
		\hat{\mathbf{S}}_{m-\ell}
		+
		\frac{1}{4}\hat{n}_{m\ell}\hat{n}_{m-\ell}
	\right)
	=
	-\frac{J}{2}
	\sum\limits_{m\sigma\sigma'}
	d_{m \ell \sigma}^\dagger
	d_{m \bar{\ell} \sigma'}^\dagger
	d_{m \ell \sigma'}
	d_{m \bar{\ell} \sigma}
	,
\end{equation}
which results in a spin-singlet ground-state
\begin{equation}
	\vert 2, E_{2_0} = 2\varepsilon_0 - \frac{J}{2}; 0, 0, 0 \rangle
	=
	\frac{1}{\sqrt{2}}\left(
		\sx{2}{2}{u}{d}{0}{0} - \sx{2}{2}{d}{u}{0}{0}
	\right)
	,
\end{equation}
a doublet of angular momentum  first excited states
\begin{equation}
	\vert 2, E_{2_1} = 2\varepsilon_0; 0, 0, 2\ell \rangle
	=
	\sx{2}{2}{2}{0}{0}{0}
	,\qquad
	\vert 2, E_{2_1} = 2\varepsilon_0;0, 0, -2\ell \rangle
	=
	\sx{2}{2}{0}{2}{0}{0}
	,
\end{equation}
and a spin-triplet of second excited states with energy $E_{2_2} = 2\varepsilon_0 + J/2$, 
\begin{equation}
	\vert 2, E_{2_2}; 1, -1, 0 \rangle
	=
	\sx{2}{2}{d}{d}{0}{0}
	,\quad
	\vert 2, E_{2_2}; 1, 0, 0 \rangle
	=
	\frac{1}{\sqrt{2}}\left(
		\sx{2}{2}{u}{d}{0}{0} + \sx{2}{2}{d}{u}{0}{0}
	\right)
	,\quad
	\vert 2, E_{2_2}; 1, 1, 0 \rangle
	=
	\sx{2}{2}{u}{u}{0}{0}
	.
\end{equation}
The one-particle dark states in Eq.~(\ref{DS_graphic})  can block transitions to the two-electron ground-state.
However, whether the blocking is effective crucially depends on the exchange energy $J$. In fact, as soon as the two-particles first excited doublet enters the transport window, interference is destroyed since transport through the doublet can occur. In our simulations we took indeed $J \simeq \Gamma =\Gamma_{\mathrm{L}} + \Gamma_\mathrm{R}$, such that the splitting of ground- and excited state is large enough to destroy interference at least partially and small enough to not see an additional excitation line appearing. Notice that in this situation the secular approximation breaks down, and the dynamics of the reduced density matrix is governed by a more general set of equations accounting also for nonsecular terms \cite{Darau2009}. 

Interestingly, for $J=0$ (or at least $J \ll \Gamma_\alpha$), the two-particle ground state, which now is a sextuplet, can form a dark state itself
\begin{equation}
	\vert 2,\mathrm{DS} \rangle
	=
	\frac{1}{2}\Bigg(
	 e^{ 2i \ell \phi_\alpha}\sx{2}{2}{2}{0}{0}{0}
	-\sx{2}{2}{u}{d}{0}{0}
	+\sx{2}{2}{d}{u}{0}{0}
	+e^{-2i \ell \phi_\alpha}\sx{2}{2}{0}{2}{0}{0}
	\Bigg)
	,
\end{equation}
which blocks transitions to the one-particle ground state at lead $\alpha$ since $\langle 1, E_{1_0}; \frac{1}{2}, \pm\frac{1}{2}, \pm\ell \vert d_{\alpha \sigma} \vert 2,\mathrm{DS} \rangle = 0$. Again we require that this dark state is not completely decoupled from the dynamics and can be reached from the left lead $\langle 1, E_{1_0}; \frac{1}{2}, \pm\frac{1}{2}, \pm\ell \vert d_{\bar{\alpha} \sigma} \vert 2,\mathrm{DS} \rangle \propto \sin(\ell(\phi_\mathrm{R}-\phi_\mathrm{L})) \neq 0$. 
\section{Impact of precession, temperature and relaxation on coherent population trapping   }
\begin{figure}
	\centering
	\includegraphics[width=\columnwidth]{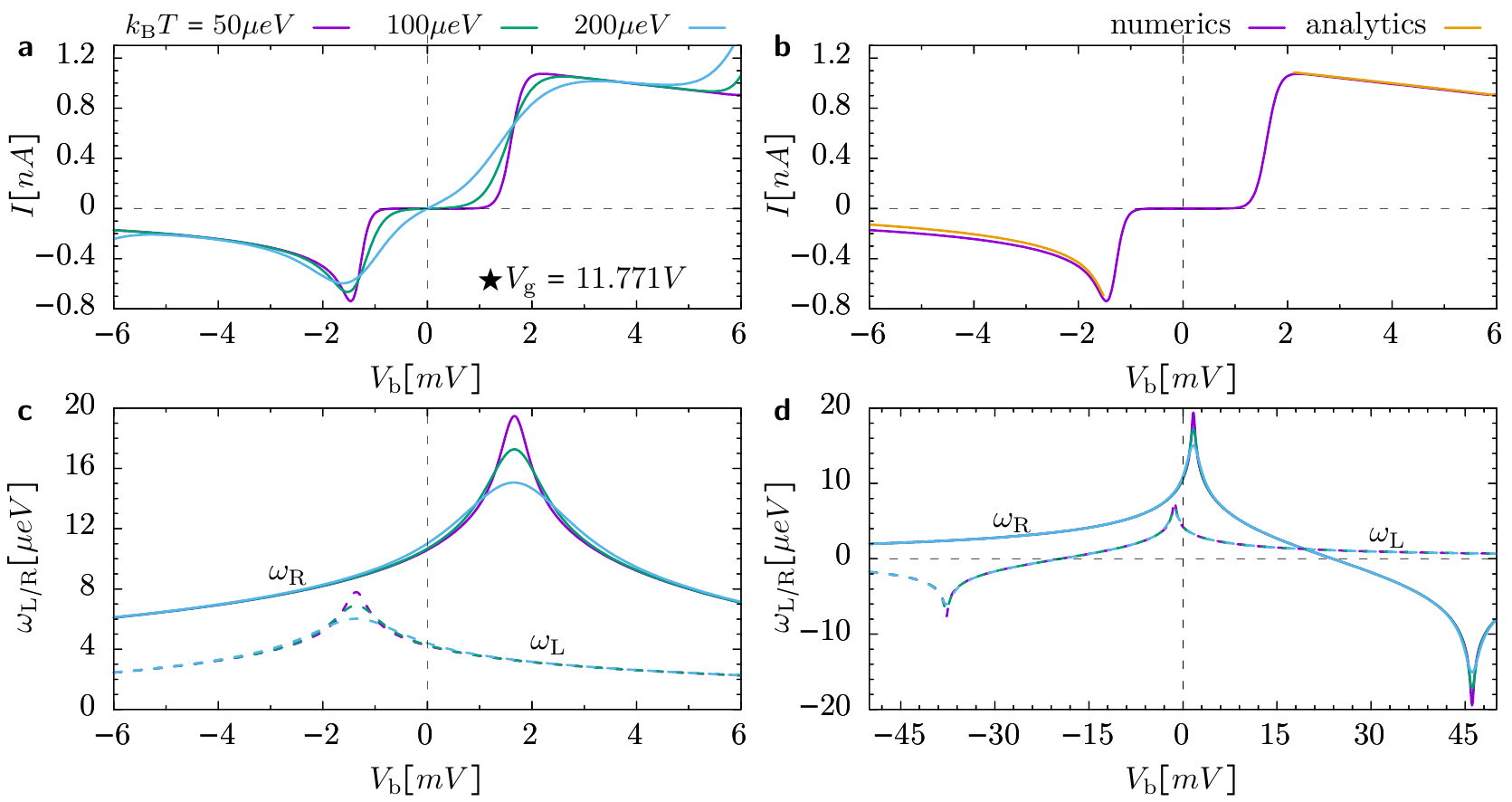}
	\caption[Bias dependence of current and precession frequencies]{
		\bs{a}
		Numerical current bias traces around the $0 \leftrightarrow 1$ resonance at $V_\mathrm{g}=11.771$V for different temperatures.
		\bs{b}
		Comparison between the numerically calculated  current and  the analytic approximation valid for vanishing relaxation. The thermal energy was set to $k_\mathrm{B}T=50\mu$eV.
		\bs{c}
		Precession frequencies s $\omega_{\mathrm{L}/\mathrm{R}}$ for the same temperatures as in (\bs{a}).
		\bs{d}
		Precession frequencies in a larger bias range. Their zeros occurs for energies $\vert e V_\mathrm{b} \vert$ of the order of the charging energy $U$.
	}
	\label{figS2}
\end{figure}
In this section we give a closer look  at the role played by the precession frequencies $\omega_{\mathrm{L}/\mathrm{R}}$, Eq.~(9) of the Methods, on the shape of the current as a function of the bias voltage. Furthermore, we investigate the role of temperature and inelastic relaxation. 
We focus on the vicinity of the $0 \leftrightarrow 1$ resonance where, as seen from the comparison in Fig.~\ref{figS2}b, the analytical expression for the current, Eq.~(3) of the main text, well reproduces the numerics.

We first start with Figs.~\ref{figS2}a, which shows the bias dependence of the numerically evaluated current for different temperatures. 
While the traditional step-like behavior of the current at positive bias for  is temperature broadened via the Fermi function, the interference peak at negative bias is not. To understand this feature, we have depicted  in Fig.~\ref{figS2}c the  precession frequencies $\omega_{\mathrm{L}/\mathrm{R}}$ in the same bias voltage range of panel (a) and for the same temperatures.  It is clear that the current changes occur in the correspondence of changes in the peaks in the precession frequencies. In particular, the temperature basically only changes the height of the resonance peaks and leaves the tails invariant. 
 
We observe that at fixed temperature the broadening  of the precession frequencies peaks is largely dominated  by the charging energy $U$, since  $\omega_{\mathrm{L}/\mathrm{R}}$ vanish when the bias becomes of the order of $U/e$, as shown in Fig.~\ref{figS2}d.
Eq.~(3) of the main text proofs that the current is dominated by a single precession frequency in the numerator, where the bias direction defines which one. If this precession frequency becomes zero, interference perfectly blocks the current. In the experiment $U$ is so large that this behavior cannot be seen; excited states enter the bias window before this value of the bias voltage is reached.

\begin{figure}
	\centering
	\includegraphics[width=\columnwidth]{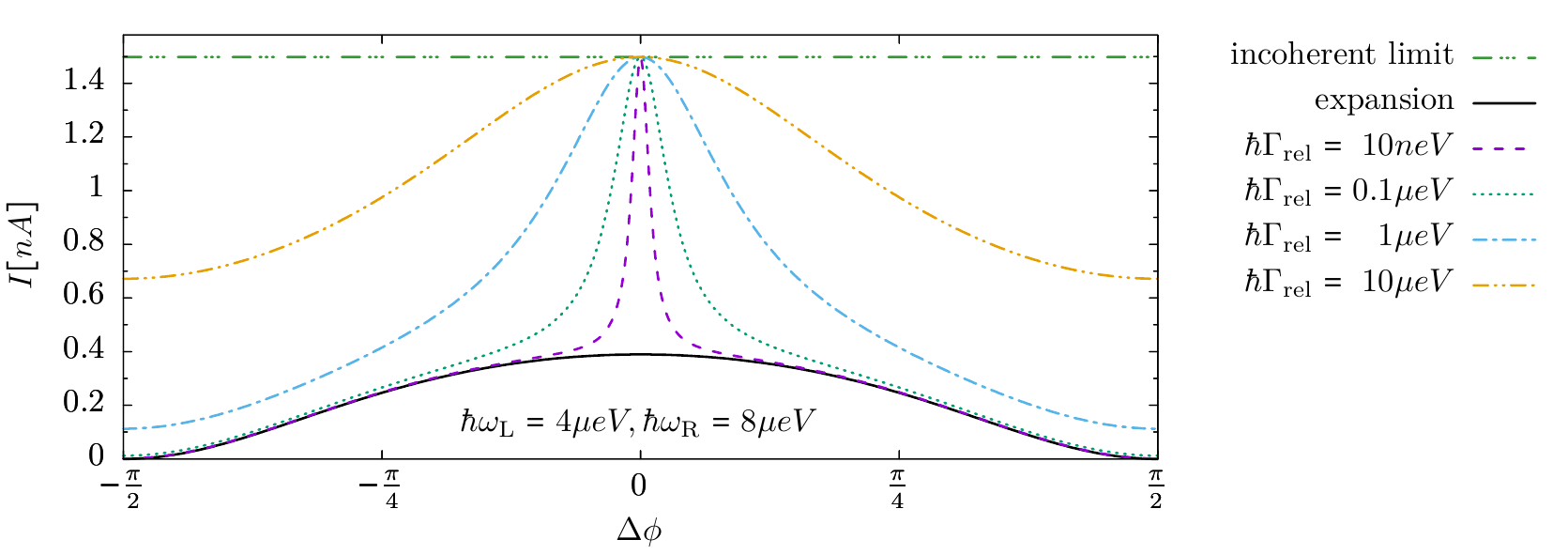}
	\caption[Dependence of the current on the phase difference between left and right contacts]{
		The numerical current is plotted for various values of  the relaxation rate $\Gamma_\mathrm{rel}$. At low relaxation the current is well approximated by the analytical expression obtained for vanishing relaxation rate (black solid line). 
		At large relaxation, the incoherent limit is approached where the  current is independent of the phase difference.
		The precession frequencies are fixed at $\hbar\omega_\mathrm{L} = 4\mu$eV and $\hbar\omega_\mathrm{R} = 8\mu$eV.
		All other parameters are the same as in Tab.~II given in the Methods section of the main manuscript.
	}
	\label{figS3}
\end{figure}

In the main manuscript we have discussed two analytic expressions for the current, valid near the $0 \leftrightarrow 1$ resonance and corresponding to the case of vanishing relaxation rate, and vanishing phase difference, respectively.
In principle Eq.~(8) from the Methods can be solved fully analytically; however  the resulting expression is very lengthy and not useful here. Therefore we display the full analytical current in Fig.~\ref{figS3} and compare it to the limiting cases.
We show the dependence of the current on the phase difference $\Delta\phi$ for various values of the relaxation rate. As an example we choose a large negative bias with precession frequencies $\hbar\omega_\mathrm{L}=3\mu$eV and $\hbar\omega_\mathrm{R} = 2\mu$eV. For low relaxation rates ($\Gamma_\mathrm{rel} \ll \Gamma_{\mathrm{L}/\mathrm{R}}$) the simple limit of Eq.~(3) is recovered at $\Delta\phi \neq 0$. This agreement is expected since in the shown bias range excited states are far-off in energy. 
At $\Delta\phi=0$ the current is always given by the incoherent limit from Eq.~(11) of the Methods section.

\section{Tunneling rate matrix}
In this section we want to show that, in general, the rate matrix defined in Eq.~(6) of the Methods,
\begin{equation}
	(\boldsymbol{\Gamma}_\alpha^m)_{\ell_z \ell_z'}(\Delta E)
	:=
	\frac{2\pi}{\hbar}
	\sum\limits_{\vec{k}}
	t^*_{\alpha \vec{k} m \ell_z}
	t_{\alpha \vec{k} m \ell_z'}
	\delta (\varepsilon_{\vec{k}} -\Delta E)
	,
\label{eq:tun_matrS}
\end{equation}
is non diagonal in the angular momentum basis. 
We then exemplify our considerations to the special case of a ring coupled to a metal, where some close analytical expressions for the rate matrix can be obtained.
\subsection{Tunneling amplitude}
We start from the tunneling Hamiltonian, whose form is given in Eq.~(5) of the Methods
\begin{equation}
	H_\mathrm{tun}
	=
	\sum\limits_{\alpha \vec{k} m \ell_z \sigma}
	t_{\alpha \vec{k} m \ell_z \sigma}
	d^\dagger_{m \ell_z \sigma}
	c_{\alpha \vec{k} \sigma}
	+
	\mathrm{h.c.}
	.
\end{equation}
The tunneling amplitude is   proportional to the overlap of a wave function of the CNT $\phi_{m \ell_z \sigma}(\vec{r}) = \langle \vec{r} \vert m \ell_z \sigma \rangle$ and one of the lead $\psi_{\alpha \vec{k} \sigma}(\vec{r}) = \langle \vec{r} \vert \alpha \vec{k} \sigma \rangle$. Explicitly, $\langle \alpha \vec{k} \sigma \vert \hat{h}  \vert m \ell_z \sigma' \rangle = t_{\alpha \vec{k} m \ell_z \sigma}\delta_{\sigma \sigma'}$,
where $\hat{h}=\frac{p^2}{2m_\mathrm{el}} +v(\vec{r})$ is the single-particle Hamiltonian of the CNT--leads complex. By decomposing the electrostatic potential into a contribution from the CNT and one from the leads, $v(\vec{r})=v_{\rm CNT}(\vec{r})+v_{\rm leads}(\vec{r})$, see the schematics in the Supplementary Figure \ref{figS4}, the tunneling amplitude  can be written as
\begin{equation}
	t_{\alpha \vec{k} m \ell_z \sigma}
	=
 	\int\limits\!\mathrm{d}\vec{r}~
	\psi_{\alpha \vec{k} \sigma}^*(\vec{r})
	\left(
		\frac{p^2}{2m_\mathrm{el}} + v(\vec{r})
	\right)
	\phi_{m \ell_z \sigma}(\vec{r})
	=	
	\langle 
		\alpha \vec{k} \sigma
	\vert \hat{h}_\mathrm{CNT} \vert
		m \ell_z \sigma
	\rangle
	+
	\underbrace{
	\langle 
		\alpha \vec{k} \sigma
	\vert v_\mathrm{leads} \vert
		m \ell_z \sigma
	\rangle
	}_{\approx 0}
	=
	\underbrace{m\varepsilon_0}_{=\varepsilon_m}
	\langle 
		\alpha \vec{k} \sigma
	\vert
		m \ell_z \sigma
	\rangle
	,
\end{equation} 
where  $\hat{h}_\mathrm{CNT}$ is the single particle part of the CNT Hamiltonian from Eq.~(4) in the Methods section. 
Since the wave functions of the CNT are much more localized than the lead ones, the contribution containing the overlap of lead and CNT wave function in the lead region  (where the potential $v_{\rm leads}$ is finite) can be neglected, yielding the simple expression 
\begin{equation}
	t_{\alpha \vec{k} m \ell_z\sigma}
	=
	\varepsilon_m
	\int \mathrm{d}\vec{r} ~
	\psi_{\alpha \vec{k} \sigma}^*(\vec{r})
	\phi_{m \ell_z \sigma}(\vec{r}).
\end{equation}
Hence, the evaluation of the tunneling amplitude requires to take a closer look at  the CNT wave functions as well as the lead wave functions in the tunneling region. 
 
We start from the latter. We assume an adiabatically smooth variation of the lead surface in the contact region, such that the lead wave functions locally factorize in a  contribution parallel to the surface and in an exponentially decaying part perpendicular to it: 
\begin{equation}
	\psi_{\alpha \vec{k} \sigma}(\vec{r})
	=
	\psi_{\alpha k_y k_z \sigma}^\parallel(y,z)
	\psi_{\alpha k_x \sigma}^\perp(x)
	=
	\frac{1}{\sqrt{L_x}}
	\psi_{\alpha k_y k_z \sigma}^\parallel(y,z) e^{-\kappa_x x}
	.
\end{equation}
Conservation of energy in the lead's potential well and in the tunneling region yields  $E_\mathrm{el}=E_\parallel -\frac{\hbar^2 \kappa_x^2}{2m_\mathrm{el}}=E_\parallel +E_b +\frac{\hbar^2 k_x^2}{2m_\mathrm{el}}$, where $E_\mathrm{el}$ is the energy of the  lead electron with respect to the vacuum. Moreover, 
the energy at the band bottom is $E_\mathrm{b}=-(E_\mathrm{F}+\phi_0$), with $E_\mathrm{F}$ the Fermi energy and $\phi_0$ the lead work function, see the Supplementary Figure \ref{figS4}. Hence,
\begin{equation}
	\kappa_x
	=
	\sqrt{
		\frac{2m_\mathrm{el}}{\hbar^2}
		\left(E_F^\alpha + \phi_0 \right)
		-k_x^2
		}
	.
\label{kappa}
\end{equation}
Thus, according to Eq.~(\ref{kappa}), the smallest values of $\kappa_x$, and hence the largest penetration  in the CNT, are obtained when $ k_x \approx k_F  $ yielding 
$\kappa_x \approx \sqrt{2m_\mathrm{el}\phi_0/\hbar^2}$. Since the total energy is bound to be $E_\mathrm{F}^\alpha$, this simultaneously implies that the longitudinal components $k_y$, $k_z$ should be vanishingly small (i.e. in the vicinity of the surface $\Gamma$--point).  
 \begin{figure*}
 	\centering
 	\includegraphics{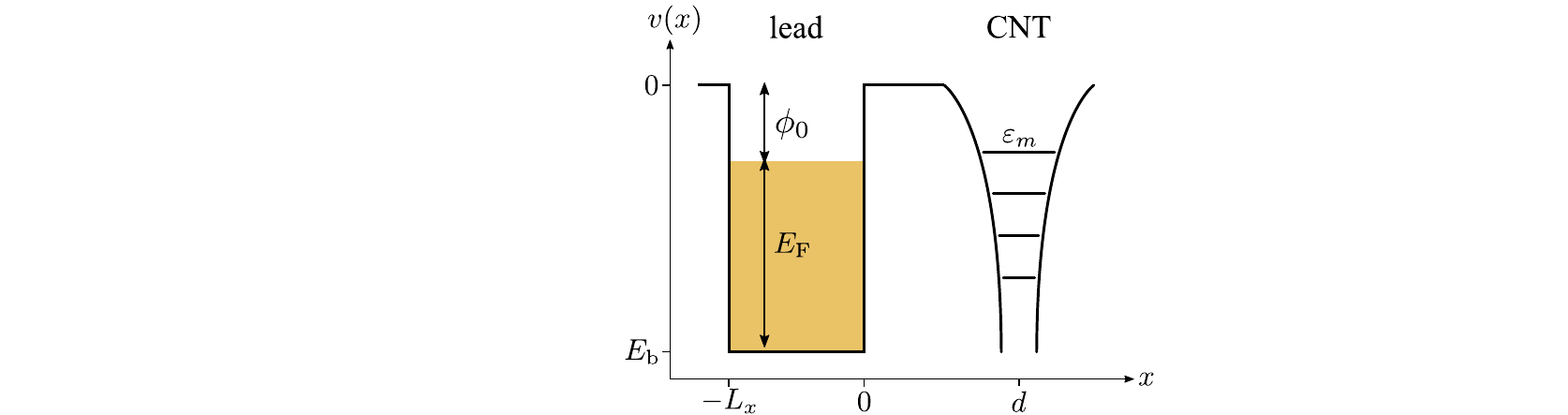}
 	\caption[Potential landscape of the CNT-lead complex]{
 		Electrostatic potential  along the tunneling direction, chosen to be along the $x$-axis.
 		In the lead, the electrons are considered to be free electrons  in the direction parallel to the surface while they experience a confinement potential in the $x$-direction.
 		$E_\mathrm{b}$ is the energy at the band bottom, $E_\mathrm{F}$ the Fermi energy and $\phi_0$ the work function. Notice that the zero of the energy has been set to the vacuum.
 		The CNT is located at a distance $d$ from the lead and features localized bound states.
 	}
 	\label{figS4}
 \end{figure*}

Regarding the  CNT wave functions, we assume that they are  well described as a linear combination of  atomic orbitals (LCAO) localized   at the atomic positions $\vec{R}_j=(X_j,Y_j,Z_j)$. 
In particular, the low energy properties are already well captured by considering a single $p$-orbital for each atomic position \cite{Dresselhaus2007}. We denote $\vert j \sigma \rangle $ such atomic state  and the associated wave function as $p_\sigma(\vec{r} -\vec{R}_j)=\langle \vec{r}\vert j \sigma \rangle $.  Hence, 
$\vert
m \ell_z \sigma	\rangle= \sum_{j \sigma} \vert
j \sigma \rangle\langle j \sigma \vert
m \ell_z \sigma	\rangle =\sum_{j\sigma} \vert j\sigma\rangle c_j(m\ell_z \sigma)$, where the LCAO coefficients $c_j(m\ell_z \sigma) \equiv \langle j \sigma \vert
m \ell_z \sigma \rangle$ have been introduced. Notice that they are chosen in such a way that the CNT wavefunctions  obey proper boundary conditions at the ends of the tube  \cite{Marganska2015}. Furthermore,  due to time reversal symmetry, it holds $c_j(m \ell_z \sigma) = c^*_j(m -\ell_z-\sigma)$. 
It follows
\begin{equation}
	t_{\alpha \vec{k} m \ell_z \sigma}
	=
	\varepsilon_m
	\sum\limits_j c_j(m\ell_z \sigma )
	\int \mathrm{d}\vec{r} ~
	\psi^*_{\alpha \vec{k} \sigma }
	(\vec{r} )p_\sigma(\vec{r} -\vec{R}_j)
	.
\end{equation}
In the {\it  absence} of spin-orbit coupling, as in our case, the spatial and spin parts factorize both for the leads as well as the CNT wave functions, yielding spin-independent coefficients $c_j(m\ell_z \sigma )=c_j(m \ell_z)$.
Similarly, the scalar product $\langle \alpha k \sigma \vert j\sigma \rangle $ becomes spin independent, yielding the final form for the tunneling amplitude
\begin{equation}
	t_{\alpha \vec{k} m \ell_z \sigma}
	=
	t_{\alpha \vec{k} m \ell_z }
	= 
	\varepsilon_m
	\sum\limits_j c_j(m\ell_z  )
	\int \mathrm{d}\vec{r} ~
	\psi^*_{\alpha \vec{k}  }(\vec{r})
	p(\vec{r} -\vec{R}_j)
	\approx 
	\varepsilon_m a
	\sum\limits_j c_j(m\ell_z  )
	\psi^*_{\alpha \vec{k}  }(\vec{R}_j )
,
\label{eq:t_final}
\end{equation} 
where in the last step we approximated the localized $p$-orbitals to Dirac-delta functions, $p(\vec{r}) = a\delta (\vec{r})$, centered at the atomic position $\vec{R}_j$. Here $a$ is a normalization factor.
The last approximation neglects the nodal plane of the $p_z$ orbitals, but it is justified by 
i) the selection of $j$ given by the lead wave function and
ii) the negligible contribution to the integral given by the CNT wave function inside the tube.

\subsection{Rate matrix of a CNT-metal complex}
We calculate the rate matrix according to Eqs.~(\ref{eq:tun_matrS}) and (\ref{eq:t_final}), i.e., in the absence of spin-orbit coupling.
We then obtain 
\begin{align}
	(\boldsymbol{\Gamma}^m_\alpha)_{\ell_z\ell_z'}(\Delta E)
	&=
	\frac{2\pi}{\hbar}
	\varepsilon^2_m 
	\vert a \vert^2
	\sum\limits_{jj'} c^*_j(m\ell_z )c_{j'}(m\ell'_z )
	\sum\limits_{\vec{k}}
	\psi_{\alpha \vec{k} }(\vec{R}_j)
	\psi^*_{\alpha \vec{k} }(\vec{R}_{j'})
	\delta(\varepsilon_{\vec{k}}-\Delta E) \nonumber
	\\
	&=
	\frac{2\pi}{\hbar}
	\frac{\vert a \vert^2}{L_x}
	\sum\limits_{jj'}
	c^*_j(m\ell_z )c_{j'}(m\ell'_z )
	\sum\limits_{\vec{k}}
	\psi^\parallel_{\alpha k_y k_z }(Y_j,Z_j)
	\psi^{\parallel*}_{\alpha k_y k_z }(Y_{j'},Z_{j'})
	e^{-\kappa_x  (X_j+X_{j'})}
	\delta(\varepsilon_{\vec{k}}-\Delta E)	.
	\label{eq:rate_full}
\end{align}
Whether the rate matrix is diagonal in the angular momentum basis, crucially depends on the geometry of the contact region. 
The exponential $e^{-\kappa_x (X_j+X_{j'})}$ in fact  selects in the sums over the atomic positions those CNT atoms closest to the leads. 
Furthermore, in the summation over the momenta $\vec{k}$, it selects  the smallest values of $\kappa_x$ compatible with the requirement that the energy of the tunneling lead electron is resonant with the CNT chemical potential $\Delta E$. As discussed in the previous subsection, this yields $k_y, k_z \approx 0$ and $\kappa_x\approx \sqrt{2m_\mathrm{el}\phi_0/\hbar^2} := \kappa_\mathrm{min}$, such that
\begin{align}
	(\boldsymbol{\Gamma}^m_\alpha)_{\ell_z\ell_z'}(\Delta E)
	&\approx
	\frac{2\pi}{\hbar}
	\varepsilon^2_m
	\frac{\vert a \vert^2}{L_x}
	\sum\limits_{jj'}
	c^*_j(m\ell_z )c_{j'}(m\ell'_z )e^{-\kappa_\mathrm{min}  (X_j+X_{j'})}
	\psi^\parallel_{\alpha k_\parallel=0 }(Y_j,Z_j)
	\psi^{\parallel*}_{\alpha k_\parallel=0 }(Y_{j'},Z_{j'})
	\sum\limits_{\vec{k}}	
	\delta(\varepsilon_{\vec{k}}-\Delta E)	.
\end{align}
Thus, this so called surface $\Gamma$-point approximation \cite{Tersoff1985} enables us to decouple the sums over $j$ and $j'$ into two independent sums. We introduce the density of states at the Fermi level $D = \sum_{\vec{k}} 	
\delta(\varepsilon_{\vec{k}} - E_\mathrm{F})$ and the tunneling coefficients 
\begin{equation}
	\tau_{\alpha}(m\ell_z)
	=
	\varepsilon_m
	\frac{a}{\sqrt{L_x}}
	\sum_{j'}
	c_{j'}(m\ell_z )
	e^{-\kappa_\mathrm{min}X_{j'}}
	\psi^{\parallel*}_{\alpha k_\parallel=0 } (Y_{j'},Z_{j'})
	,
\end{equation}
yielding 
\begin{align}
	(\boldsymbol{\Gamma}^m_\alpha)_{\ell_z\ell_z'}(\Delta E)
	&=
	\frac{2\pi}{\hbar}D\tau^*_{\alpha}(m\ell_z)\tau_{\alpha}(m\ell'_z)
	.
\end{align}
In the surface $\Gamma$-point approximation, the wave function $\psi^\parallel(Y_j,Z_j)$ is independent of $k_y$ and $k_z$ and hence real.
Furthermore, the LCAO coefficients are related by time-reversal symmetry, $c_j(m,\ell_z)=c^*_j(m,-\ell_z)$, yielding the result $\tau_{\alpha}(m\ell_z)=\tau^*_{\alpha}(m-\ell_z)$. Accounting for this symmetry we finally obtain the final form for the rate matrix used in the main text 
\begin{equation}
	(\boldsymbol{\Gamma}^m_\alpha)_{\ell_z\ell_z'}(\Delta E)
	=
	\Gamma^m_\alpha
	e^{i\phi_\alpha^m(\ell_z -\ell'_z)}
	, \quad 
	\Gamma_\alpha^m
	=\frac{2\pi}{\hbar}
	D\vert \tau_{\alpha}(m\ell_z)\vert^2
	, \quad
	\phi_\alpha^m
	:=
	arg\{\tau_\alpha (m\ell)\}
	.
\label{eq:rate_final}
\end{equation}

The result in Equation~(\ref{eq:rate_final}) strongly relies on the surface $\Gamma$-point approximation, which allows one to decouple the double sum over the atomic positions $j$ and $j'$. In the next subsection we have explored the consequences of keeping a finite contribution for the parallel momenta $k_y$ and $k_z$ on the example of a  ring of $N$ carbon atoms coupled in three different ways to a metal.
As we shall see, if the ring is lying flat on the substrate, such that all atoms are equally distant from the lead, the rate matrix becomes diagonal, see Supplementary Figure \ref{figS5}a.
The result in Eq.~(\ref{eq:rate_final}) is in contrast recovered when the ring is  orthogonal to the substrate, in a way that tunneling is dominated by only one  closest atom, Supplementary Figure \ref{figS5}b. When two atoms are equally close to the surface, as for the case in figure \ref{figS5}c, the rate matrix is off-diagonal, but the modulus of the diagonal elements is smaller than that of the diagonal ones. 
 
From this we conclude that in the CNT case, where only few atoms are close to the leads, the rate matrix is not diagonal.
How good the simple form Eq. (\ref{eq:rate_final}) describes the experiment, depends on various factors, among which tube's chirality. In the case of our experiment, we consider CNTs of the zig-zag class, which at the tube's end have non vanishing weights only for atoms of a given sublattice \cite{Marganska2015}.    
Thus, if at the left end only $A$ atoms have non vanishing LCAO coefficients, this implies that the neighboring $B$ atoms are not tunneling coupled, hence effectively achieving the situation described in the simple example in Fig.~(\ref{figS5})b.
At the right lead, the same considerations apply upon exchange of the role of atoms  $A$ and $B$.

\subsection{Rate matrix of a ring-metal complex }
In this subsection we calculate the rate matrix for a ring of $N$ carbon atoms with radius $R$. To study the effects of the orientation of the ring with respect to the surface, we study three different configurations. In the  first one, the ring is lying parallel to the surface of the lead, at a distance $d$ from it,  as shown in the Supplementary  Figure~\ref{figS5}a; in the second one it is standing on the $x-y$ plane perpendicular to the lead plane like shown in Figure~\ref{figS5}b; in the third one it is also standing but now in a way that two atoms are equally distant from the lead, as shown in Figure~\ref{figS5}c.
The rate matrix for the ring follows easily from the general expression Eq.~(\ref{eq:rate_full}) upon dropping the shell index $m$. Similar to the CNT, the ring has a C$_N$ symmetry and its single particle eigenstates can be classified in terms of angular momentum, $\ell_z$, and  spin, $\sigma$, degrees of freedom. The LCAO coefficients follow from the diagonalization of the ring Hamiltonian, and have the form $c_j(\ell_z)=\langle j \vert \ell_z \rangle = \frac{1}{\sqrt{N}} e^{i\frac{2\pi}{N} j\ell_z}$.
Eq.~(\ref{eq:rate_full}) yields then for the ring the rate matrix 
\begin{align}
	(\boldsymbol{\Gamma}_\alpha)_{\ell_z \ell_z'}(\Delta E)
	&=
	\frac{2\pi}{\hbar}
	\frac{\varepsilon^2}{N}
	\vert a \vert^2
	\sum\limits_{jj'}
	e^{-i\frac{2\pi}{N}(j\ell_z-j'\ell_z')}	
	\sum\limits_{\vec{k}}
	\psi_{\alpha \vec{k} }(\vec{R}_j)
	\psi^*_{\alpha \vec{k} }(\vec{R}_{j'})
	\delta(\varepsilon_{\vec{k}}-\Delta E)
	\\
	&=
	\frac{2\pi}{\hbar}
	\frac{\varepsilon^2}{N}
	\frac{\vert a \vert^2 }{L_x}
	\sum\limits_{jj'}
	e^{-i\frac{2\pi}{N}(j\ell_z-j'\ell_z')}
	\sum\limits_{\vec{k}}
	\psi^\parallel_{\alpha \vec{k_\parallel} }(\vec{R}_j)
	\psi^{\parallel*}_{\alpha \vec{k_\parallel} }(\vec{R}_{j'})
	e^{	-\kappa_x (X_j+X_{j'})}
	\delta(\varepsilon_{\vec{k}}-\Delta E)
	.
\end{align}
To further simplify the calculation we consider in the following a plane wave behavior for the parallel wave function, $\psi^\parallel_{\alpha \vec{k_\parallel} }(\vec{R})=\frac{1}{\sqrt{S}}e^{i\vec{k_\parallel}\cdot \vec{R}}$, with $S$ a normalization constant.
\\ 

\noindent\bs{Case a}\\
Let us consider the first case where  the ring is lying planar on top of the lead at a distance $X_j=d$. The rate matrix then reads
\begin{equation}
	(\boldsymbol{\Gamma}_\alpha)_{\ell_z\ell_z'}(\Delta E)
	=
	\frac{2\pi}{\hbar}
	\frac{\mathcal{C} \varepsilon^2}{N}
	\sum\limits_{jj'}
	e^{-i\frac{2\pi}{N} (j\ell_z - j'\ell_z')}
	\sum\limits_{\vec{k}}
	e^{i\vec{k}_\parallel \cdot (\vec{R}_j - \vec{R}_{j'})}
	e^{	-2\kappa_x d}
	\delta(\epsilon_{\vec{k}}-\Delta E)
	,
\end{equation}
where $\mathcal{C} =: \vert a \vert^2/(SL_x)$.
It is convenient to express $j'=j'-j+j:=-\Delta j +j $ and to observe that $\vert \vec{R}_j - \vec{R}_{j'}\vert := R_{\Delta j}$ only depends on the relative distance $\Delta j$ but not on the position $j$.
This suggests to transform the sum over momentum into an integral, and to express this integral in cylindrical coordinates $k_y, k_z \to \varphi, k_\parallel$. This results in
\begin{equation}
	(\boldsymbol{\Gamma}_\alpha)_{\ell_z\ell_z'}(\Delta E)
	=
	\frac{2\pi}{\hbar}
	\frac{\mathcal{C} \varepsilon^2}{N}
	\overbrace{
		\sum\limits_{j}
		e^{-i\frac{2\pi}{N} j(\ell_z - \ell_z')}
	}^{N\delta_{\ell_z\ell_z'}}
	\sum\limits_{\Delta j}
	e^{-i\frac{2\pi}{N} \Delta j \ell_z'} 
	\int \mathrm{d}\varphi
	\int \mathrm{d}k_\parallel
	\int \mathrm{d}k_x
	e^{i k_\parallel R_{\Delta j} \cos \varphi}
	k_\parallel
	e^{-2\kappa_x d}
	\delta(\varepsilon_{\vec{k}}-\Delta E)
	.
\end{equation}
The integration over the angle $\varphi $ results in a real function of $k_\parallel$; similarly, in the sum over $\Delta j$ for each finite positive  $\Delta j$ there is a negative counterpart. This leaves us with the   diagonal and real rate matrix
\begin{equation}	
\label{eq:Gamma_a}
	\boldsymbol{\Gamma}_\alpha
	=
	\Gamma_\alpha\begin{pmatrix}
		1 & 0 \\
		0 & 1
	\end{pmatrix}
	,
\end{equation}
which therefore does not support dark states. The absolute value of the diagonal parts is not important in this consideration.
For simplicity and homogeneity with the other cases we have kept the $\delta$ approximation for the $p_z$ functions. The description of a CNT ``slice'' would rather require to consider the orbital structure of the radially distributed $p$ orbitals.
Eq.~(\ref{eq:Gamma_a}) is obtained, though, also out of more fundamental symmetry arguments:
\begin{align}
\label{eq:Gamma_a_sym}
	(\boldsymbol{\Gamma}_\alpha)_{\ell_z \ell_z'}
	&=
	\frac{2\pi}{\hbar}
	\varepsilon_m^2
	\sum\limits_{\vec{k}}
	\langle m \ell_z \vert \alpha \vec{k} \rangle
	\!
	\langle \alpha \vec{k} \vert m \ell_z' \rangle
	=
	\frac{2\pi}{\hbar}
	\varepsilon_m^2
	\sum\limits_{\vec{k}}
	\langle m \ell_z \vert
	\hat{C}^\dagger_N \hat{C}_N
	\vert	
	\alpha \vec{k} \rangle
	\!
	\langle \alpha \vec{k} \vert
	\hat{C}^\dagger_N \hat{C}_N
	\vert
	m \ell_z' \rangle
	\nonumber \\
	&=
	\frac{2\pi}{\hbar}
	\varepsilon_m^2
	e^{i\frac{2\pi}{N}(\ell_z' - \ell_z)}
	\sum\limits_{\vec{k}}
	\langle m \ell_z \vert \alpha \vec{k} \rangle
	\!
	\langle \alpha \vec{k} \vert m \ell_z' \rangle
	=
	e^{i\frac{2\pi}{N}(\ell_z' - \ell_z)}
	(\boldsymbol{\Gamma}_\alpha)_{\ell_z \ell_z'}
	,
\end{align}
where $\hat{C}_N$ is the rotation of $2\pi/N$ around the $x$ axis and the isotropy of the leads is assumed.
Eq.~(\ref{eq:Gamma_a_sym}) implies $\boldsymbol{\Gamma}_\alpha$ is diagonal. The form in Eq.~(\ref{eq:Gamma_a}) follows by requiring time reversal symmetry.
\begin{figure}
	\centering
	\includegraphics{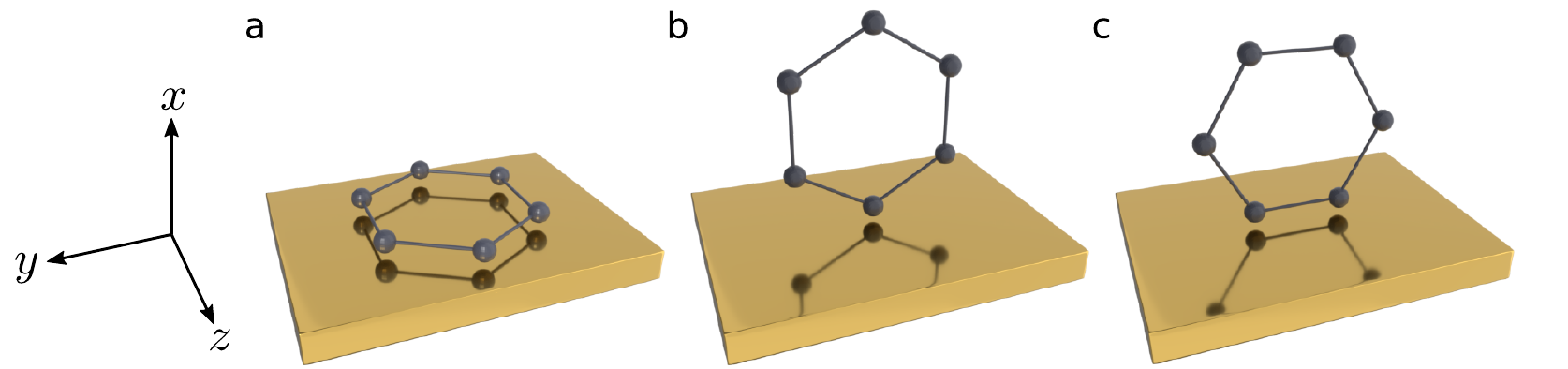}
	\caption[Tunneling configurations for a ring-lead complex]{
		\bs{a}
		The ring lies flat on the lead surface. All atoms of the ring are equidistant from the surface. 
		\bs{b}
		The ring is standing perpendicular to the lead  with only one carbon atom closest to the surface.
		\bs{c}
		Similar to the previous case, the ring is standing perpendicular to the surface  but rotated such that two carbon atoms are equidistant from  the lead.
	}
	\label{figS5}
\end{figure}
\\ 

\noindent\bs{Case b}\\
In the second case the result is quite different. The rate matrix for the standing ring is
\begin{equation}
	(\boldsymbol{\Gamma}_\alpha)_{\ell_z\ell_z'}(\Delta E)
	=
	\frac{2\pi}{\hbar}
	\frac{\mathcal{C}\varepsilon^2}{N}
	\sum\limits_{jj'}
	e^{-i\frac{2\pi}{N} (j\ell_z - j'\ell_z')}
	\sum\limits_{\vec{k}}
	e^{
		ik_y (Y_j - Y_{j'})
		-\kappa_x (X_j + X_{j'})
	}
	\delta(\varepsilon_{\vec{k}}-\Delta E)
	,
\end{equation}
where $\mathcal{C} = \vert a\vert^2/(SL_x)$.
One can see immediately that the trick used in the previous case $a$ does not work here, due to the dependence on $j$ $(j')$  of the variables $X_j$ ($X_j'$).
An estimation of $
\kappa_x 
= 
\sqrt{
	\frac{2m_\mathrm{el}}{\hbar^2}
	\left(E_F^\alpha + \phi_0^\alpha \right) 
	- 
	k_x^2
} 
\geq 
\sqrt{
	\frac{2m_\mathrm{el}}{\hbar^2}\phi_0^\alpha
} 
= 
\mathcal{O}(\text{\AA}^{-1})
$ for typical work functions $\phi_0=\mathcal{O}(\mathrm{eV})$ tells that the contribution to the rate matrix shrinks by one order of magnitude for a distance of $1$\AA{} of the atom to the lead surface.
This suggests that perfect local tunneling to the atom closest to the lead $j = j' = J$ at distance $X_J=d$ is a good approximation. We then obtain
\begin{equation}
	(\boldsymbol{\Gamma}_\alpha)_{\ell_z\ell_z'}
	=
	\frac{2\pi}{\hbar}
	\frac{\mathcal{C}\varepsilon^2}{N}
	e^{-iJ\frac{2\pi}{N} (\ell_z - \ell_z')}
	\sum\limits_{\vec{k}}
	e^{-2\kappa^\alpha_x d}
	\delta(\epsilon_{\vec{k}}-\Delta E)
	\qquad \Rightarrow \qquad
	\boldsymbol{\Gamma}_\alpha
	=
	\Gamma_\alpha \begin{pmatrix}
		1 & e^{2i\phi_\alpha} \\
		e^{-2i\phi_\alpha} & 1
	\end{pmatrix}
	.
\end{equation}
Thus the single atom contact yields a rate matrix with maximal coherence, like in the surface $\Gamma$-point approximation discussed in the previous subsection.
%
\\

\noindent \bs{Case c}\\
In the third case the ring is rotated in a way that tunneling can occur through two atoms $J$ and $J'$ which are both in contact with the lead ($X_J=X_{J'}=d$). The rate matrix reads
\begin{equation}
	(\boldsymbol{\Gamma}_\alpha)_{\ell_z\ell_z'}(\Delta E)
	=
	\frac{2\pi}{\hbar}
	\frac{\mathcal{C}\varepsilon^2}{N}
	\sum\limits_{jj' \in (J,J')}
	e^{-i\frac{2\pi}{N} (j\ell_z - j'\ell_z')}
	\sum\limits_{\vec{k}}
	e^{ik_y (Y_j - Y_{j'}) - 2\kappa_x d}
	\delta(\epsilon_{\vec{k}}-\Delta E)
	.
\end{equation}
The diagonal and off-diagonal elements of the rate matrix can be simplified to
\begin{align}
	(\boldsymbol{\Gamma}_\alpha)_{\ell\ell}(\Delta E)
	&=
	\frac{4\pi}{\hbar}
	\frac{\mathcal{C}\varepsilon^2}{N}
	\sum\limits_{\vec{k}}
	\left[
		1 + 
		\cos\left( \frac{2\pi}{N}\ell(J-J') \right)
		\cos\left( k_y \Delta Y \right)
	\right]
	e^{-2\kappa_x d}
	\delta(\epsilon_{\vec{k}}-\Delta E)
	,
	\\
	(\boldsymbol{\Gamma}_\alpha)_{\ell-\ell}(\Delta E)
	&=
	\frac{4\pi}{\hbar}
	\frac{\mathcal{C}\varepsilon^2}{N}
	e^{-i\frac{2\pi}{N}\ell (J+J')}
	\sum\limits_{\vec{k}} 
	\left[
		\cos\left(
			\frac{2\pi}{N}\ell(J-J')
		\right)
		+
		\cos\left(
			k_y \Delta Y
		\right)
	\right]
	e^{-2\kappa_x d}
	\delta(\epsilon_{\vec{k}}-\Delta E)
	,
\end{align}
with $\Delta Y = Y_J - Y_{J'}$. We used the fact that the sum over $\vec{k}$ is isotropic and therefore $\sin(k_y\Delta Y)\rightarrow0$.
One can see directly that for $\Delta Y=0$ the result of case \bs{b} is recovered. For $\Delta Y \neq 0$ the amplitude of the off-diagonal terms in $\boldsymbol{\Gamma}_\alpha$ is smaller than the diagonal values since in general $1 + \cos x \cos y - \cos x - \cos y = (\cos x -1)(\cos y -1) \geq 0$. We obtain a rate matrix intermediate to cases \bs{a} and \bs{b}
\begin{equation}
	\boldsymbol{\Gamma}_\alpha
	=
	\Gamma_\alpha\begin{pmatrix}
		1 & h e^{2i\phi_\alpha} \\
		h e^{-2i\phi_\alpha} & 1
	\end{pmatrix}
	,\quad
	0 \leq h \leq 1
	.
\end{equation}

\begin{table}
\begin{minipage}{0.5\linewidth}
\centering
\begin{tabular}{ l | c | c | c | c }
\multicolumn{5}{c}{$N=0$} \\
\hline
\hline
$\Delta E$
 & $S$
  & $S_z$
   & $L_z$
    & Eigenstate \\
\hline
$0$
 & $0$
  & $0$
   & $0$
    & $\ssx{2}{2}{0}{0}{0}{0}$ \\
\hline
\multirow{13}{*}{$\varepsilon_0-\dfrac{J}{2}$}
 & \multirow{2}{*}{$0$}
  & \multirow{2}{*}{$0$}
   & \multirow{2}{*}{$0$}
    & $\frac{1}{\sqrt{2}}\left(\ssx{1}{2}{-1}{0}{0}{0}-\ssx{-1}{2}{1}{0}{0}{0}\right)$ \\
\cline{5-5}
 &
  & 
   & 
   	& $\frac{1}{\sqrt{2}}\left(\ssx{2}{1}{-1}{0}{0}{0}-\ssx{2}{-1}{1}{0}{0}{0}\right)$ \\
\cline{2-5}
 & \multirow{3}{*}{$1$}
  & $1$
   & \multirow{3}{*}{$0$}
   	& $\ssx{1}{2}{1}{0}{0}{0}$ \\
\cline{3-3}
\cline{5-5}
 &
  & $0$
   & 
    & $\frac{1}{\sqrt{2}}\left(\ssx{1}{2}{-1}{0}{0}{0}+\ssx{-1}{2}{1}{0}{0}{0}\right)$ \\
\cline{3-3}
\cline{5-5}
 & 
  & $-1$
   & 
    & $\ssx{-1}{2}{-1}{0}{0}{0}$ \\
\cline{2-5}
 & $0$
  & $0$
   & $2\ell$
    & $\frac{1}{\sqrt{2}}\left(\ssx{2}{-1}{1}{0}{0}{0}-\ssx{2}{1}{-1}{0}{0}{0}\right)$ \\
\cline{2-5}
 & \multirow{3}{*}{$1$}
  & $1$
   & \multirow{3}{*}{$2\ell$}
    & $\ssx{2}{1}{1}{0}{0}{0}$ \\
\cline{3-3}
\cline{5-5}
 & 
  & $0$
   & 
    & $\frac{1}{\sqrt{2}}\left(\ssx{2}{1}{-1}{0}{0}{0}+\ssx{2}{-1}{1}{0}{0}{0}\right)$ \\
\cline{3-3}
\cline{5-5}
 & 
  & $-1$
   & 
    & $\ssx{2}{-1}{-1}{0}{0}{0}$ \\
\cline{2-5}
 & $0$
  & $0$
   & $-2\ell$
    & $\frac{1}{\sqrt{2}}\left(\ssx{-1}{2}{0}{1}{0}{0}+\ssx{1}{2}{0}{-1}{0}{0}\right)$ \\
\cline{2-5}
 & \multirow{3}{*}{$1$}
  & $1$
   & \multirow{3}{*}{$-2\ell$}
    & $\ssx{1}{2}{1}{0}{0}{0}$ \\
\cline{3-3}
\cline{5-5}
 & 
  & $0$
   & 
    & $\frac{1}{\sqrt{2}}\left(\ssx{1}{2}{-1}{0}{0}{0}+\ssx{-1}{2}{1}{0}{0}{0}\right)$ \\
\cline{3-3}
\cline{5-5}
 & 
  & $-1$
   & 
    & $\ssx{-1}{2}{-1}{0}{0}{0}$ \\
\hline
\hline
\end{tabular}
\end{minipage}\begin{minipage}{0.5\linewidth}
\centering
\begin{tabular}{ l | c | c | c | c }
\multicolumn{5}{c}{$N=1$} \\
\hline
\hline
$\Delta E$
 & $S$
  & $S_z$
   & $L_z$
    & Eigenstate \\
\hline
\multirow{4}{*}{$0$}
 & \multirow{2}{*}{$\dfrac{1}{2}$}
  & $\frac{1}{2}$
   & \multirow{2}{*}{$\ell$}
    & $\ssx{2}{2}{1}{0}{0}{0}$ \\
\cline{3-3}
\cline{5-5}
 & 
  & $-\frac{1}{2}$
   & 
    & $\ssx{2}{2}{-1}{0}{0}{0}$ \\
\cline{2-5}
 & \multirow{2}{*}{$\dfrac{1}{2}$}
  & $\frac{1}{2}$
   & \multirow{2}{*}{$-\ell$}
    & $\ssx{2}{2}{0}{1}{0}{0}$ \\
\cline{3-3}
\cline{5-5}
 & 
  & $-\frac{1}{2}$
   & 
    & $\ssx{2}{2}{0}{-1}{0}{0}$ \\
\hline
\multirow{4}{*}{$\varepsilon_0-J$}
 & \multirow{2}{*}{$\dfrac{1}{2}$}
  & $\frac{1}{2}$
   & \multirow{2}{*}{$\ell$}
    & $\frac{1}{\sqrt{2}}\left(\ssx{2}{1}{1}{-1}{0}{0}-\ssx{2}{1}{-1}{1}{0}{0}\right)$ \\
\cline{3-3}
\cline{5-5}
 & 
  & $-\frac{1}{2}$
   & 
    & $\frac{1}{\sqrt{2}}\left(\ssx{2}{-1}{1}{-1}{0}{0}-\ssx{2}{-1}{-1}{1}{0}{0}\right)$ \\
\cline{2-5}
 & \multirow{2}{*}{$\dfrac{1}{2}$}
  & $\frac{1}{2}$
   & \multirow{2}{*}{$-\ell$}
    & $\frac{1}{\sqrt{2}}\left(\ssx{1}{2}{1}{-1}{0}{0}-\ssx{1}{2}{-1}{1}{0}{0}\right)$ \\
\cline{3-3}
\cline{5-5}
 & 
  & $-\frac{1}{2}$
   & 
    & $\frac{1}{\sqrt{2}}\left(\ssx{-1}{2}{1}{-1}{0}{0}-\ssx{-1}{2}{-1}{1}{0}{0}\right)$ \\
\hline
\multirow{8}{*}{$\varepsilon_0-\dfrac{J}{2}$}
 & \multirow{2}{*}{$\dfrac{1}{2}$}
  & $\frac{1}{2}$
   & \multirow{2}{*}{$3\ell$}
    & $\ssx{2}{1}{2}{0}{0}{0}$ \\
\cline{3-3}
\cline{5-5}
 & 
  & $-\dfrac{1}{2}$
   & 
    & $\ssx{2}{-1}{2}{0}{0}{0}$ \\
\cline{2-5}
 & \multirow{2}{*}{$\dfrac{1}{2}$}
  & $\frac{1}{2}$
   & \multirow{2}{*}{$\ell$}
    & $\ssx{1}{2}{2}{0}{0}{0}$ \\
\cline{3-3}
\cline{5-5}
 & 
  & $-\frac{1}{2}$
   & 
    & $\ssx{-1}{2}{2}{0}{0}{0}$ \\
\cline{2-5}
 & \multirow{2}{*}{$\dfrac{1}{2}$}
  & $\frac{1}{2}$
   & \multirow{2}{*}{$-\ell$}
    & $\ssx{2}{1}{0}{2}{0}{0}$ \\
\cline{3-3}
\cline{5-5}
 & 
  & $-\frac{1}{2}$
   & 
    & $\ssx{2}{-1}{0}{2}{0}{0}$ \\
\cline{2-5}
 & \multirow{2}{*}{$\dfrac{1}{2}$}
  & $\frac{1}{2}$
   & \multirow{2}{*}{$-3\ell$}
    & $\ssx{1}{2}{0}{2}{0}{0}$ \\
\cline{3-3}
\cline{5-5}
 & 
  & $-\frac{1}{2}$
   & 
    & $\ssx{-1}{2}{0}{2}{0}{0}$ \\
\hline
\hline
\end{tabular}
\end{minipage}
\caption[Eigenstates of a CNT for N=0,1]{First few eigenstates of a three shell CNT Hamiltonian with $N=0$ (left) and $N=1$ (right) electrons above the reference state. $\Delta E$ is the energy above the ground states with energies $E_{0}$ and $E_{1} = E_{0}+\varepsilon_0+U/2$.}
\label{tab:CNT_states_01}
\end{table}

\begin{table}
\begin{minipage}{0.5\textwidth}
\centering
\begin{tabular}{ l | c | c | c | c }
\multicolumn{5}{c}{$N=1$} \\
\hline
\hline
$\Delta E$
 & $S$
  & $S_z$
   & $L_z$
    & Eigenstate \\
\hline
\multirow{12}{*}{$\varepsilon_0$}
 & \multirow{2}{*}{$\dfrac{1}{2}$}
  & $\frac{1}{2}$
   & \multirow{2}{*}{$\ell$}
    & $\ssx{2}{2}{0}{0}{1}{0}$,
    	$\frac{1}{\sqrt{6}}\left(
    	\ssx{2}{1}{-1}{1}{0}{0}
    	+\ssx{2}{1}{1}{-1}{0}{0}
    	-2\ssx{2}{-1}{1}{1}{0}{0}
    	\right)$ \\
\cline{3-3}
\cline{5-5}
 & 
  & $-\frac{1}{2}$
   & 
    & $\ssx{2}{2}{0}{0}{-1}{0}$,
    	$\frac{1}{\sqrt{6}}\left(
    	\ssx{2}{-1}{1}{-1}{0}{0}
    	+\ssx{2}{-1}{-1}{1}{0}{0}
    	-2\ssx{2}{1}{-1}{-1}{0}{0}
    	\right)$ \\
\cline{2-5}
 & \multirow{2}{*}{$\dfrac{1}{2}$}
  & $\frac{1}{2}$
   & \multirow{2}{*}{$-\ell$}
    & $\ssx{2}{2}{0}{0}{0}{1}$,
    	$\frac{1}{\sqrt{6}}\left(
    	\ssx{1}{2}{1}{-1}{0}{0}
    	+\ssx{1}{2}{-1}{1}{0}{0}
    	-2\ssx{-1}{2}{1}{1}{0}{0}
    	\right)$ \\
\cline{3-3}
\cline{5-5}
 & 
  & $-\frac{1}{2}$
   & 
    & $\ssx{2}{2}{0}{0}{0}{-1}$,
    	$\frac{1}{\sqrt{6}}\left(
    	\ssx{-1}{2}{1}{-1}{0}{0}
    	+\ssx{-1}{2}{-1}{1}{0}{0}
    	-2\ssx{1}{2}{-1}{-1}{0}{0}
    	\right)$ \\
\cline{2-5}
 & \multirow{4}{*}{$\dfrac{3}{2}$}
  & $\frac{3}{2}$
   & \multirow{4}{*}{$\ell$}
    & $\ssx{2}{1}{1}{1}{0}{0}$ \\
\cline{3-3}
\cline{5-5}
 & 
  & $\frac{1}{2}$
   & 
    & $\frac{1}{\sqrt{3}}\left(
    	\ssx{2}{1}{-1}{1}{0}{0}
    	+\ssx{2}{1}{1}{-1}{0}{0}
    	+\ssx{2}{-1}{1}{1}{0}{0}
    	\right)$ \\
\cline{3-3}
\cline{5-5}
 &
  & $-\frac{1}{2}$
   &
    & $\frac{1}{\sqrt{3}}\left(
    	\ssx{2}{1}{-1}{-1}{0}{0}
    	+\ssx{2}{-1}{-1}{1}{0}{0}
    	+\ssx{2}{-1}{1}{-1}{0}{0}
    	\right)$ \\
\cline{3-3}
\cline{5-5}
 & 
  & $-\frac{3}{2}$
   & 
    & $\ssx{2}{-1}{-1}{-1}{0}{0}$ \\
\cline{2-5}
 & \multirow{4}{*}{$\dfrac{3}{2}$}
  & $\frac{3}{2}$
   & \multirow{4}{*}{$-\ell$}
    & $\ssx{1}{2}{1}{1}{0}{0}$ \\
\cline{3-3}
\cline{5-5}
 & 
  & $\frac{1}{2}$
   & 
    & $\frac{1}{\sqrt{3}}\left(
    	\ssx{1}{2}{1}{-1}{0}{0}
    	+\ssx{1}{2}{-1}{1}{0}{0}
    	+\ssx{-1}{2}{1}{1}{0}{0}
    	\right)$ \\
\cline{3-3}
\cline{5-5}
 &
  & $-\frac{1}{2}$
   &
    & $\frac{1}{\sqrt{3}}\left(
    	\ssx{1}{2}{-1}{-1}{0}{0}
    	+\ssx{-1}{2}{1}{-1}{0}{0}
    	+\ssx{-1}{2}{-1}{1}{0}{0}
    	\right)$ \\
\cline{3-3}
\cline{5-5}
 & 
  & $-\frac{3}{2}$
   & 
    & $\ssx{-1}{2}{-1}{-1}{0}{0}$ \\
\hline
\hline
\end{tabular}
\end{minipage}\begin{minipage}{0.5\textwidth}
\centering
\begin{tabular}{ l | c | c | c | c }
\multicolumn{5}{c}{$N=2$} \\
\hline
\hline
$\Delta E$
 & $S$
  & $S_z$
   & $L_z$
    & Eigenstate \\
\hline
$0$
 & $0$
  & $0$
   & $0$
    & $\frac{1}{\sqrt{2}}\left(
    	\ssx{2}{2}{1}{-1}{0}{0}
    	-\ssx{2}{2}{-1}{1}{0}{0}
    	\right)$ \\
\hline
\multirow{2}{*}{$\dfrac{J}{2}$}
 & \multirow{2}{*}{$0$}
  & \multirow{2}{*}{$0$}
   & $2\ell$
    & $\ssx{2}{2}{2}{0}{0}{0}$ \\
\cline{4-5}
 & 
  & 
   & $-2\ell$
    & $\ssx{2}{2}{0}{2}{0}{0}$ \\
\hline
\multirow{3}{*}{$J$}
 & \multirow{3}{*}{$1$}
  & $1$
   & \multirow{3}{*}{$0$}
    & $\ssx{2}{2}{1}{1}{0}{0}$ \\
\cline{3-3}
\cline{5-5}
 & 
  & $0$
   & 
    & $\frac{1}{\sqrt{2}}\left(
    	\ssx{2}{2}{1}{-1}{0}{0}
    	+\ssx{2}{2}{-1}{1}{0}{0}
    	\right)$ \\
\cline{3-3}
\cline{5-5}
 & 
  & $-1$
   & 
    & $\ssx{2}{2}{-1}{-1}{0}{0}$ \\
\hline
\multirow{14}{*}{$\begin{matrix}\varepsilon_0 \\ + \\ \dfrac{J}{2}\end{matrix}$}
 & $0$
  & $0$
   & $2\ell$
    & $\frac{1}{\sqrt{2}}\left(
    	\ssx{2}{-1}{2}{1}{0}{0}
    	-\ssx{2}{1}{2}{-1}{0}{0}
    	\right)$,%
    	$\frac{1}{\sqrt{2}}\left(
    	\ssx{2}{2}{1}{0}{-1}{0}
    	-\ssx{2}{2}{-1}{0}{1}{0}
    	\right)$ \\
\cline{2-5}
 & \multirow{3}{*}{$1$}
  & $1$
   & \multirow{3}{*}{$2\ell$}
    & $\ssx{2}{1}{2}{1}{0}{0}$, $\ssx{2}{2}{1}{0}{1}{0}$ \\
\cline{3-3}
\cline{5-5}
 & 
  & $0$
   & 
    & $\frac{1}{\sqrt{2}}\left(
    	\ssx{2}{-1}{2}{1}{0}{0}
    	+\ssx{2}{1}{2}{-1}{0}{0}
    	\right)$,%
    	$\frac{1}{\sqrt{2}}\left(
    	\ssx{2}{2}{-1}{0}{1}{0}
    	+\ssx{2}{2}{1}{0}{-1}{0}
    	\right)$ \\
\cline{3-3}
\cline{5-5}
 & 
  & $-1$
   & 
    & $\ssx{2}{-1}{2}{-1}{0}{0}$, $\ssx{2}{2}{-1}{0}{-1}{0}$ \\
\cline{2-5}
 & \multirow{2}{*}{$0$}
  & \multirow{2}{*}{$0$}
   & \multirow{2}{*}{$0$}
    & $\frac{1}{\sqrt{2}}\left(
    	\ssx{2}{-1}{1}{2}{0}{0}
    	-\ssx{2}{1}{-1}{2}{0}{0}
    	\right)$,%
    	$\frac{1}{\sqrt{2}}\left(
    	\ssx{-1}{2}{2}{1}{0}{0}
    	-\ssx{1}{2}{2}{-1}{0}{0}
    	\right)$, \\
 & 
  &
   & 
    & $\frac{1}{\sqrt{2}}\left(
    	\ssx{2}{2}{0}{-1}{1}{0}
    	-\ssx{2}{2}{0}{1}{-1}{0}
    	\right)$,%
    	$\frac{1}{\sqrt{2}}\left(
    	\ssx{2}{2}{-1}{0}{0}{1}
    	-\ssx{2}{2}{1}{0}{0}{-1}
    	\right)$ \\
\cline{2-5}
 & \multirow{4}{*}{$1$}
  & $1$
   & \multirow{4}{*}{$0$}
    & $\ssx{2}{1}{1}{2}{0}{0}$,
    	$\ssx{1}{2}{2}{1}{0}{0}$,
    	$\ssx{2}{2}{0}{1}{1}{0}$,
    	$\ssx{2}{2}{1}{0}{0}{1}$ \\
\cline{3-3}
\cline{5-5}
 & 
  & \multirow{2}{*}{$0$}
   & 
    & $\frac{1}{\sqrt{2}}\left(
    	\ssx{2}{-1}{1}{2}{0}{0}
    	+\ssx{2}{1}{-1}{2}{0}{0}
    	\right)$,%
    	$\frac{1}{\sqrt{2}}\left(
    	\ssx{-1}{2}{2}{1}{0}{0}
    	+\ssx{1}{2}{2}{-1}{0}{0}
    	\right)$, \\
 & 
  &
   & 
    & $\frac{1}{\sqrt{2}}\left(
    	\ssx{2}{2}{0}{-1}{1}{0}
    	+\ssx{2}{2}{0}{1}{-1}{0}
    	\right)$,%
    	$\frac{1}{\sqrt{2}}\left(
    	\ssx{2}{2}{-1}{0}{0}{1}
    	+\ssx{2}{2}{1}{0}{0}{-1}
    	\right)$ \\
\cline{3-3}
\cline{5-5}
 &
  & $0$
   & 
    & $\ssx{2}{-1}{-1}{2}{0}{0}$,
    	$\ssx{-1}{2}{2}{-1}{0}{0}$,
    	$\ssx{2}{2}{0}{-1}{-1}{0}$,
    	$\ssx{2}{2}{-1}{0}{0}{-1}$ \\
\cline{2-5}
 & $0$
  & $0$
   & $-2\ell$
    & $\frac{1}{\sqrt{2}}\left(
    	\ssx{-1}{2}{1}{2}{0}{0}
    	-\ssx{1}{2}{-1}{2}{0}{0}
    	\right)$,%
    	$\frac{1}{\sqrt{2}}\left(
    	\ssx{2}{2}{-1}{0}{1}{0}
    	-\ssx{2}{2}{1}{0}{-1}{0}
    	\right)$ \\
\cline{2-5}
 & \multirow{3}{*}{$1$}
  & $1$
   & \multirow{3}{*}{$-2\ell$}
    & $\ssx{1}{2}{1}{2}{0}{0}$,
    	$\ssx{2}{2}{2}{0}{0}{0}$ \\
\cline{3-3}
\cline{5-5}
 & 
  & $0$
   & 
    & $\frac{1}{\sqrt{2}}\left(
    	\ssx{-1}{2}{1}{2}{0}{0}
    	+\ssx{1}{2}{-1}{2}{0}{0}
    	\right)$,%
    	$\frac{1}{\sqrt{2}}\left(
    	\ssx{2}{2}{-1}{0}{1}{0}
    	+\ssx{2}{2}{1}{0}{-1}{0}
    	\right)$ \\
\cline{3-3}
\cline{5-5}
 & 
  & $-1$
   & 
    & $\ssx{-1}{2}{-1}{2}{0}{0}$,
    	$\ssx{2}{2}{-1}{0}{-1}{0}$ \\
\hline
\hline
\end{tabular}
\end{minipage}
\caption[Eigenstates of a CNT for N=1,2]{Next few eigenstates of a three shell CNT Hamiltonian with $N=1$ (left) electrons above the reference state and the first few eigenstates with $N=2$ (right). $\Delta E$ is the energy above the ground states with energies $E_{1} = E_{0}+\varepsilon_0+U/2$ and $E_{2} = E_{0}+2\varepsilon_0+2U -J/2$.}
\label{tab:CNT_states_12}
\end{table}

\begin{table}
\begin{minipage}{0.5\linewidth}
\centering
\begin{tabular}{ l | c | c | c | c }
\multicolumn{5}{c}{$N=3$} \\
\hline
\hline
$\Delta E$
 & $S$
  & $S_z$
   & $L_z$
    & Eigenstate \\
\hline
\multirow{4}{*}{$0$}
 & \multirow{2}{*}{$\dfrac{1}{2}$}
  & $\frac{1}{2}$
   & \multirow{2}{*}{$\ell$}
    & $\ssx{2}{2}{2}{1}{0}{0}$ \\
\cline{3-3}
\cline{5-5}
 & 
  & $-\frac{1}{2}$
   & 
    & $\ssx{2}{2}{2}{-1}{0}{0}$ \\
\cline{2-5}
 & \multirow{2}{*}{$\dfrac{1}{2}$}
  & $\frac{1}{2}$
   & \multirow{2}{*}{$-\ell$}
    & $\ssx{2}{2}{1}{2}{0}{0}$ \\
\cline{3-3}
\cline{5-5}
 & 
  & $-\frac{1}{2}$
   & 
    & $\ssx{2}{2}{-1}{2}{0}{0}$ \\
\hline
\multirow{4}{*}{$\varepsilon_0-J$}
 & \multirow{2}{*}{$\dfrac{1}{2}$}
  & $\frac{1}{2}$
   & \multirow{2}{*}{$\ell$}
    & $\frac{1}{\sqrt{2}}\left(
    	\ssx{2}{2}{1}{-1}{1}{0}
    	-\ssx{2}{2}{-1}{1}{1}{0}
    	\right)$ \\
\cline{3-3}
\cline{5-5}
 & 
  & $-\frac{1}{2}$
   & 
    & $\frac{1}{\sqrt{2}}\left(
    	\ssx{2}{2}{1}{-1}{-1}{0}
    	-\ssx{2}{2}{-1}{1}{-1}{0}
    	\right)$ \\
\cline{2-5}
 & \multirow{2}{*}{$\dfrac{1}{2}$}
  & $\frac{1}{2}$
   & \multirow{2}{*}{$-\ell$}
    & $\frac{1}{\sqrt{2}}\left(
    	\ssx{2}{2}{1}{-1}{0}{1}
    	-\ssx{2}{2}{-1}{1}{0}{1}
    	\right)$ \\
\cline{3-3}
\cline{5-5}
 & 
  & $-\frac{1}{2}$
   & 
    & $\frac{1}{\sqrt{2}}\left(
    	\ssx{2}{2}{1}{-1}{0}{-1}
    	-\ssx{2}{2}{-1}{1}{0}{-1}
    	\right)$ \\
\hline
\multirow{8}{*}{$\varepsilon_0-\dfrac{J}{2}$}
 & \multirow{2}{*}{$\dfrac{1}{2}$}
  & $\frac{1}{2}$
   & \multirow{2}{*}{$3\ell$}
    & $\ssx{2}{2}{2}{0}{1}{0}$ \\
\cline{3-3}
\cline{5-5}
 & 
  & $-\frac{1}{2}$
   & 
    & $\ssx{2}{2}{2}{0}{-1}{0}$ \\
\cline{2-5}
 & \multirow{2}{*}{$\dfrac{1}{2}$}
  & $\frac{1}{2}$
   & \multirow{2}{*}{$\ell$}
    & $\ssx{2}{2}{2}{0}{0}{1}$ \\
\cline{3-3}
\cline{5-5}
 & 
  & $-\frac{1}{2}$
   & 
    & $\ssx{2}{2}{2}{0}{0}{-1}$ \\
\cline{2-5}
 & \multirow{2}{*}{$\dfrac{1}{2}$}
  & $\frac{1}{2}$
   & \multirow{2}{*}{$-\ell$}
    & $\ssx{2}{2}{0}{2}{1}{0}$ \\
\cline{3-3}
\cline{5-5}
 & 
  & $-\frac{1}{2}$
   & 
    & $\ssx{2}{2}{0}{2}{-1}{0}$ \\
\cline{2-5}
 & \multirow{2}{*}{$\dfrac{1}{2}$}
  & $\frac{1}{2}$
   & \multirow{2}{*}{$-3\ell$}
    & $\ssx{2}{2}{0}{2}{0}{1}$ \\
\cline{3-3}
\cline{5-5}
 & 
  & $-\frac{1}{2}$
   & 
    & $\ssx{2}{2}{0}{2}{0}{-1}$ \\
\hline
\hline
\end{tabular}
\end{minipage}\begin{minipage}{0.5\linewidth}
\centering
\begin{tabular}{ l | c | c | c | c }
\multicolumn{5}{c}{$N=3$} \\
\hline
\hline
$\Delta E$
 & $S$
  & $S_z$
   & $L_z$
    & Eigenstate \\
\hline
\multirow{12}{*}{$\varepsilon_0$}
 & \multirow{2}{*}{$\dfrac{1}{2}$}
  & $\frac{1}{2}$
   & \multirow{2}{*}{$\ell$}
    & $\ssx{2}{1}{2}{2}{0}{0}$,
    	$\frac{1}{\sqrt{6}}\left(
    	\ssx{2}{2}{1}{-1}{1}{0}
    	+\ssx{2}{2}{-1}{1}{1}{0}
    	-2\ssx{2}{2}{1}{1}{-1}{0}
    	\right)$ \\
\cline{3-3}
\cline{5-5}
 & 
  & $-\frac{1}{2}$
   & 
    & $\ssx{2}{-1}{2}{2}{0}{0}$,
    	$\frac{1}{\sqrt{6}}\left(
    	\ssx{2}{2}{-1}{1}{-1}{0}
    	+\ssx{2}{2}{1}{-1}{-1}{0}
    	-2\ssx{2}{2}{-1}{-1}{1}{0}
    	\right)$ \\
\cline{2-5}
 & \multirow{2}{*}{$\dfrac{1}{2}$}
  & $\frac{1}{2}$
   & \multirow{2}{*}{$-\ell$}
    & $\ssx{1}{2}{2}{2}{0}{0}$,
    	$\frac{1}{\sqrt{6}}\left(
    	\ssx{2}{2}{-1}{1}{0}{1}
    	+\ssx{2}{2}{1}{-1}{0}{1}
    	-2\ssx{2}{2}{1}{1}{0}{-1}
    	\right)$ \\
\cline{3-3}
\cline{5-5}
 & 
  & $-\frac{1}{2}$
   & 
    & $\ssx{-1}{2}{2}{2}{0}{0}$,
    	$\frac{1}{\sqrt{6}}\left(
    	\ssx{2}{2}{1}{-1}{0}{-1}
    	+\ssx{2}{2}{-1}{1}{0}{-1}
    	-2\ssx{2}{2}{-1}{-1}{0}{1}
    	\right)$ \\
\cline{2-5}
 & \multirow{4}{*}{$\dfrac{3}{2}$}
  & $\frac{3}{2}$
   & \multirow{4}{*}{$\ell$}
    & $\ssx{2}{2}{1}{1}{1}{0}$ \\
\cline{3-3}
\cline{5-5}
 & 
  & $\frac{1}{2}$
   & 
    & $\frac{1}{\sqrt{3}}\left(
    	\ssx{2}{2}{-1}{1}{1}{0}
    	+\ssx{2}{2}{1}{-1}{1}{0}
    	+\ssx{2}{2}{1}{1}{-1}{0}
    	\right)$ \\
\cline{3-3}
\cline{5-5}
 &
  & $-\frac{1}{2}$
   &
    & $\frac{1}{\sqrt{3}}\left(
    	\ssx{2}{2}{1}{-1}{-1}{0}
    	+\ssx{2}{2}{-1}{1}{-1}{0}
    	+\ssx{2}{2}{-1}{-1}{1}{0}
    	\right)$ \\
\cline{3-3}
\cline{5-5}
 & 
  & $-\frac{3}{2}$
   & 
    & $\ssx{2}{2}{-1}{-1}{-1}{0}$ \\
\cline{2-5}
 & \multirow{4}{*}{$\dfrac{3}{2}$}
  & $\frac{3}{2}$
   & \multirow{4}{*}{$-\ell$}
    & $\ssx{2}{2}{1}{1}{0}{1}$ \\
\cline{3-3}
\cline{5-5}
 & 
  & $\frac{1}{2}$
   & 
    & $\frac{1}{\sqrt{3}}\left(
    	\ssx{2}{2}{-1}{1}{0}{1}
    	+\ssx{2}{2}{1}{-1}{0}{1}
    	+\ssx{2}{2}{1}{1}{0}{-1}
    	\right)$ \\
\cline{3-3}
\cline{5-5}
 &
  & $-\frac{1}{2}$
   &
    & $\frac{1}{\sqrt{3}}\left(
    	\ssx{2}{2}{1}{-1}{0}{-1}
    	+\ssx{2}{2}{-1}{1}{0}{-1}
    	+\ssx{2}{2}{-1}{-1}{0}{1}
    	\right)$ \\
\cline{3-3}
\cline{5-5}
 & 
  & $-\frac{3}{2}$
   & 
    & $\ssx{2}{2}{-1}{-1}{0}{-1}$ \\
\hline
\hline
\end{tabular}
\end{minipage}
\caption[Eigenstates of a CNT for N=3]{First few eigenstates of a three shell CNT Hamiltonian with $N=3$ electrons above the reference state. $\Delta E$ is the energy above the ground state with energy $E_{3} = E_{0} + 3\varepsilon_0 + 9U/2 + J/2$.}
\label{tab:CNT_states_3}
\end{table}

\clearpage

%

\end{document}